%% file: localization.tex
\documentclass[11pt]{article}
\usepackage[a4paper,bindingoffset=0.2in,%
            left=0.9in,right=0.9in,top=1in,bottom=1in,%
            footskip=.25in]{geometry}
\usepackage[pdftex]{graphicx}
\usepackage{amsfonts,amsmath,amsthm,amscd,amssymb,array}
\usepackage{subcaption}
\usepackage[printonlyused]{acronym}
\newcommand{\mbf}{\mathbf}
\newcommand{\mbb}{\mathbb}

\newcommand{\mrm}{\mathrm}

\DeclareMathOperator*{\gl}{\gtrless}

\begin{document} 
\input{acronym_def}

\title{Learning the Localization Function: Machine Learning Approach to Fingerprinting Localization}

\author{Linchen Xiao, Arash Behboodi and Rudolf Mathar\\
\\
Institute for Theoretical Information Technology, RWTH Aachen University
}

\date{}

\maketitle

\begin{abstract}
Considered as a data-driven approach, \acp{FPS} enjoy huge popularity due to their good performance and minimal environment information requirement. This papers addresses applications of artificial intelligence to solve two problems in \ac{RSSI} based \ac{FPS}, first the cumbersome training database construction  and second the extrapolation of fingerprinting algorithm for similar buildings with slight environmental changes. After a concise overview of deep learning design techniques, two main techniques widely used in deep learning are exploited for the above mentioned issues namely data augmentation and transfer learning. We train a multi-layer neural network that learns the mapping from the observations to the locations. A data augmentation method is proposed to increase the training database size based on the structure of \ac{RSSI} measurements and hence reducing effectively the amount of training data. Then it is shown experimentally how a model trained for a particular building can be transfered to a similar one by  fine tuning with significantly smaller training numbers. The paper implicitly discusses the new guidelines to consider about deep learning designs when they are employed in a new application context.
\end{abstract}

\section{Introduction}
Precise location of \textit{things} in indoor environments is an essential information for future wireless networks and services. Significant research has been conducted during recent years on indoor localization. In plethora of indoor localization algorithms, RF based approaches are particularly interesting given their technological accessibility. 
However the propagation in indoor environments follow complex models that should account for various sources of attenuation and deflection. Given the dynamics of indoor environments, the propagation models should also be constantly updated with changes in the propagation space. Therefore the algorithms that rely on explicit propagation models for localization  
require significant environmental awareness and continuous manual update. 
 
Fingerprinting-based methods, on the other hand, are not model-dependent. They are data-driven approaches working on the assumption that there are certain RF features capable of identifying a location uniquely and stably. The algorithm collect these features at different locations and constructs fingerprint for each point. The fingerprint collection can be done for only finite number of points in the space and it is the most time-consuming part of the algorithm. These pairs of fingerprints and locations are organized into a training database. The localization boils down to finding  the location corresponding to a new observation. Fingerprinting algorithms can usually be built on top of available infrastructures such as WiFi networks and their model-independence and minimal infrastructure requirement make them a very attractive choice for fast indoor localization deployment. As their drawbacks, the training database should be updated and sometimes built anew when the environment changes and consequently so are the fingerprints. This is more troubling given the time consuming nature of data collection. The same problem exists when one considers two similar environments for examples two floors of a single building with a very similar structure. Intuitively the similar environments share many structural features that might potentially facilitate the fingerprinting process. In this paper, the goal is to look at the fingerprinting localization algorithms from machine learning point of view and show how these issues can be addressed using the modern learning architectures. Not only utilization of these learning architectures improves upon the classic algorithms in term of accuracy but the techniques like transfer learning and data augmentation can be borrowed from machine learning to accelerate new training base creation. 

There are abundance of works on \ac{RF} based indoor localizations  \cite{medina_ultrasound_2013, brassart_localization_2000, erol-kantarci_survey_2011, amundson_survey_2009,Seco2009} and particularly fingerprinting solutions \cite{Milioris2014,honkavirta2009comparative}. Machine learning approaches have received attention in \cite{milioris2011low}, neural network-based approach in \cite{Laoudias2009}, $K$-means algorithm in ~\cite{Bai2013} and prediction-based training methods in \cite{Steiner2011}. There are many works addressing the theoretical issues around fingerprinting methods, for example the effect of number of \acp{AP} on the localization performance~\cite{Machaj2010}. The general theoretical framework for fingerprinting algorithms is presented in~\cite{Kaemarungsi2004,Kaemarungsi2005,wen_fundamental_2015,behboodi17hypothesis,behboodi_mathematical_2016,BehboodiIPIN2017} where the interaction of many design parameters are discussed such as  radio propagation parameters, the training grid, the number of measurements and interference. The issue around the database construction and scalability has been discussed in~\cite{Ding2013}. There have been some researches which already applied deep learning to indoor localization problems. In \cite{wang2017csi}, the authors propose a fingerprint construction using neural networks based on measured \acp{CSI}.  In \cite{nowicki2016low}, \ac{RSSI} values are used to solve floor classification problem. In \cite{Linchenetal2017}, the authors introduce a neural network for mapping the fingerprints to the locations directly.

\subsection{Summary of Contributions}
In this paper, the main motivations are to address some challenges in fingerprinting localization solutions using modern artificial intelligence such as \acp{DNN} and to examine whether  new insights are required for designing \acp{DNN} when they are used in a new application context. We will consider first fingerprinting algorithms as a data-driven approach from the perspective of data science. The databased organization is discussed from the point of view of existing techniques. Deep architectures like autoencoders are successful in extracting useful features from image datasets. It is investigated whether they can be as effective for \ac{RSSI}-based databases.
Next assuming that the localization algorithm is nothing but a function that maps certain observations to a location, a \ac{DNN}, that is a multi-layer neural network, is  trained to learn this function. On the other hand, it is shown that certain interesting features of \acp{DNN}, namely transfer learning and data augmentation, can be used to address the problems of extrapolation across similar environments and faster training construction. 

The rest of the paper is structured as follows. Section \ref{sec:architecture} presents the general overview of fingerprinting algorithms particularly from data analysis point of view.  Section \ref{sec:SVM} discusses fingerprinting as a classification problem. In Section \ref{sec:DeepLearning}, deep learning architectures are proposed  for fingerprinting and genera design guidelines are discussed further in Sections \ref{sec:design} and \ref{sec:transfer}. Section \ref{sec:Numerical} verifies the effectiveness of the design through experimental and simulation based performance evaluations.

\section{Architecture of Fingerprinting Algorithms}
\label{sec:architecture}
The localization space is given by the set $\mathcal{D}\subset \mathbb{R}^d$. The goal of localization is to infer the position of a target node placed at $\mathbf u$ in the region $\mathcal D$ based on the observations $\mathbf S_{\mathbf u}$, usually a vector in Euclidean space. Therefore the task of localization can be understood as the problem learning the mapping of the observation $\mathbf S$ to the location $\mathbf u$.  Fingerprinting algorithms are data-driven approaches where the localization task utilizes the data gathered from the environment as the basis for location inference. In fingerprinting algorithm, a set of true location-observation pairs $(\mathbf v,\mathbf S_{\mathbf v})$ is known and the localization function is learned using the set mapping new observations to a position. The central task of fingerprinting localization is to learn the localization function denoted by $\Phi$.  The building blocks of fingerprinting algorithms are well known. We briefly overview the essentials. The algorithm can be roughly decomposed into raw signal feature collection, fingerprint creation, pattern matching and post-processing  \cite{lemic2014experimental}. 
As it will be explained below, these phases correspond to different phases of a supervised learning algorithm.

\begin{figure}[h]
\centering
\includegraphics[height = 3cm,width=9cm]{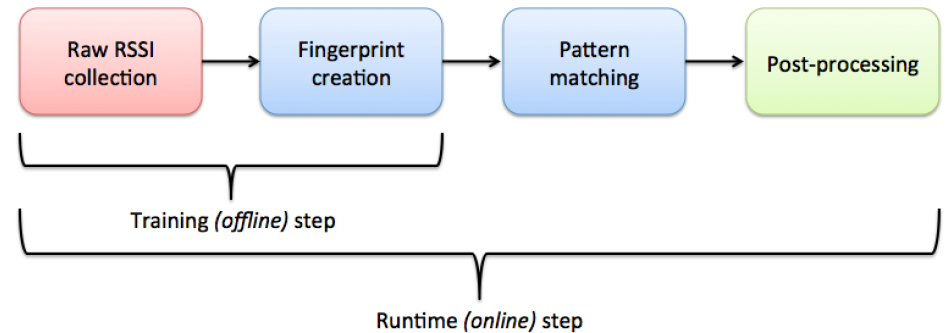}
\caption{\scriptsize Phases of fingerprinting algorithm \cite{lemic2014experimental}}
\end{figure}

\subsection{Raw Feature Collection}
As any data-driven learning algorithm, the fingerprinting algorithm requires the collection of data. In this phase, the algorithm selects a grid of points, denoted by $\Lambda$, in the localization space $\mathcal{D}$. Some choices of the grid include square, hexagonal and random grid. At each point $\mathbf v$ in this grid $\Lambda$, also called training points, 
some observations are made that pertain to the location information. For RF-based localization, a signal feature is measured, mostly multiple times to compensate the transient effect of propagation environment such as fading and shadowing. The feature is chosen such that it contains adequate information about the location. The sufficient condition for the signal feature to distinguish training points is discussed in \cite{behboodi17hypothesis,behboodi_mathematical_2016}. It states simply that the probabilistic descriptions of the feature at different locations should be enough distinct measured in terms of \ac{KL} divergence. In this work, \ac{RSSI} values of multiple anchors are used as the signal feature. The measurements are labeled with the location of the training points and therefore construct a database of labeled data.

\subsection{Fingerprint Creation and Pre-Processing}
Fingerprint creation is considered a kind of pre-processing for the data that is essentially heterogeneous. In data analysis applications, it is in general essential to pre-process the data for better presentation, compression and cleaning of the data. There are various guidelines for preparing the data for further analysis \cite{kandel_wrangler_2011, wickham_tidy_2014}. Our data preparation follows the tiny data paradigm developed in \cite{wickham_tidy_2014}. The tiny data paradigm is closely based on Codd’s relational algebra \cite{codd_relational_1990}. The tidy data provides a standard way for data preparation and cleaning. The idea is to organize the data such that to each column corresponds a variable and to each row corresponds a measurement which in this case is just a training point. An example is shown in Table \ref{FPdatabase}.

\begin{table}[]
\centering
\caption{Fingerprinting Database}
\label{FPdatabase}
\begin{tabular}{|c|c|c|c|c|}
\hline
Location 		& Anchor ID	 	& Measurement 1		& Measurement 2	& $\dots$ \\ \hline\hline
$\mathbf v=(x,y)$	& $\mrm{AP}_\mrm{ID}$	& $S_1$ 		& $S_2$		&$\dots$ \\ \hline
\end{tabular}
\end{table}	

In RF based fingerprinting solutions, given the volatile nature of wireless environments, measurements from different anchors differ in their numbers and scaling. At a location, one might not be able to get the same number of measurements from anchors due to their different signal strengths and incurred packet loss. Therefore the numbers of visible access points and the corresponding measurements differ from location to location. In this work, the training set consists of rows of different size corresponding to each training location. The row contains the position of the training point $\mathbf v$ and the anchor ID and the corresponding measurements. In the light of what we discussed above, each row might have different length. It will be discussed later how to employ different data augmentation techniques to increase the size of training data and construct the rows of same length. At the end, one constructs the training dataset of form $(\mathbf v,\mathbf X_{\mathbf v})$ from the raw observations $(\mathbf v,\mathbf S_{\mathbf v})$ and $\mathbf X_{\mathbf v}$ is called the fingerprint of the point $\mathbf v$. 

For RF-based fingerprinting, a simple and popular procedure\cite{honkavirta2009comparative} for building a homogeneous database out of heterogeneous data is to compute the  average value of \ac{RSSI} values obtained from each access point.  The averaged values are arranged into a vector and hence the rows of final database consists of the training location and the vector of averaged \ac{RSSI} value. This vector is the fingerprint of the point $\mathbf v\in\Lambda$. There are other ways to create the fingerprint including, to name a few, \ac{RSSI} quantiles or fitting Multivariate Gaussian distributions to \acp{RSSI}. In this work, the averaging method is used as the benchmark for comparison with other fingerprinting algorithm. In general, once the data is prepared for further analysis, the fingerprint of the point $\mathbf v$ is implicitly understood as the rows corresponding to $\mathbf v$ in the database.

\subsection{Pattern Matching and Post-processing}

Once the training dataset is built, a function should be learned mapping the new observations to positions in $\mathcal D$. In conventional fingerprinting algorithms, the function of pattern matching is to capture the similarity between fingerprints of training points and the fingerprint of test points. Namely, the goal is to find the most similar pairs of test point and training point in the fingerprint space and then use the location information of the training points to estimate the test point in the location space.
 
One of the most well known algorithms for pattern matching employs \ac{ED} to measure the similarity between fingerprints given as
\begin{equation}
    d(\mathbf{X}_{\mathbf u},\mathbf{X}_{\mathbf v}) = \|\mathbf{X}_{\mathbf u}-\mathbf{X}_{\mathbf v}\|_2.
\end{equation}
With a new test point to be estimated, the Euclidean distance has to be calculated between the fingerprints $\mbf X_{\mbf v}$ of all training points $\mathbf v$ in training grid $\Lambda$ and the fingerprint of test point $\mbf X$. The training point with smallest Euclidean distance from test point would be the best candidate:
\begin{equation}
 \hat{\mbf u}=\arg\underset{\mbf v\in \Lambda}{\min}\{d(\mbf{X}_{\mbf v},\mbf{X})\}.   
\end{equation}
In general, the function $d$ can be any kernel function. This process is completed with post-processing methods such as $k-$nearest neighbors is the last step of traditional fingerprint algorithms. In the previous step, the $k$ closest fingerprints in the training set are chosen and the final location is obtained by the linear combination of the $k$ corresponding locations. The number $k$ and the weights of final linear combination as parameters of learning algorithm are chosen during the matching process. 

Indeed, this approach is nothing but $k$-nearest neighbor classifier. The problem of learning the localization function $\Phi$ is a supervised learning problem. Depending on the particular problem at hand, it can be a classification or regression problem. Regression-based localization function aims at giving the estimated location while classification-based localization function identifies the room or the area in which the target node is placed. 
In this paper, we consider both classification and regression-based approach and address the design challenges of learning algorithms in this context.
\subsection{System Model}
As mentioned above, \ac{RSSI}-based fingerprinting algorithms are considered in this work. At each point, a number of \ac{RSSI} measurements is taken from visible \acp{AP}. The number of visible \acp{AP} and measured \acp{RSSI} might differ from point to point. 

\section{Localization as Classification}
\label{sec:SVM}
\begin{figure}[th]
  \centering
 \includegraphics[width=0.7\textwidth]{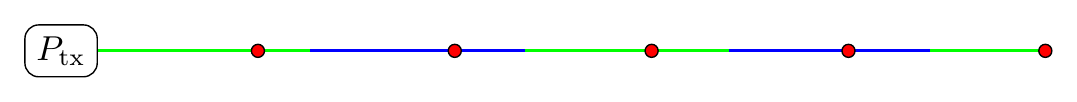}
  \caption{Partition of the localization space based on Euclidean distance of \acp{RSSI} }
    \label{fig:EuclbasedFP}
\end{figure}

A challenging feature of \ac{RSSI} based fingerprinting is that the mapping from locations to their corresponding measured \acp{RSSI} is not a uniformly continuous mapping. In other words, for points extremely close to an \ac{AP}, small change in the distance leads to arbitrary big difference between measurements. Moreover the mapping is far from being an isometric mapping. This means that the proximity of \ac{RSSI} values do not necessarily correspond to the geometric proximity. To see that consider a simple line localization where an \ac{AP} is placed at the origin. Training points are placed at $(2k,0)$ for $k\in\mathbb{N}_{>0}$ and \ac{RSSI} values are the measured received power at each point.  If the Euclidean distance is used as pattern matching metric for \ac{RSSI}-based fingerprinting, then the localization space is partitioned into intervals containing the training points. Each region contains all the points whose fingerprints are the closest in  Euclidean distance to the fingerprint of the training point in the interval. However the intervals are not symmetric around the training point as one might expect. This can be seen in Fig. \ref{fig:EuclbasedFP}. Although in this particular problem, this situation can be avoided by using the inverse \ac{RSSI} values, it will not be useful in a general indoor environment with multiple anchors and complex propagation structure. Therefore in general Euclidean distance based fingerprinting leads to uneven partition of space around the training points. In other words the closest fingerprint does not translate into the closest training point. One way to circumvent the problem is to use additional data that are not as precise as training points but contain general proximity information. In other words, these new measurements  are labeled by the closest training point \textit{in geometric sense} instead of the precise location. In this sense, the localization is considered as a classification problem. The training grid $\Lambda$ divides the localization space into different regions geometrically and the goal of the localization algorithm is to determine the geometric region corresponding to a test point by looking at \ac{RSSI} measurements. Therefore each class corresponds to one of these regions. This formulation makes it possible to utilize the opportunistic measurements obtained by crowd-sourcing approaches. Among classification techniques, \acp{SVM} are used here as the classification algorithm. 

\subsection{Support Vector Machine Algorithm}
\ac{SVM} \cite{cortes1995support} is a binary classifier introduced by Vapnik and Chervonenkis in 1963. \ac{SVM} aims at finding a linear classifier, i.e., a hyperplane which maximizes the margin between two classes. It has extensions to non-linear classifiers and non-separable data too.  
 \begin{figure}[h]
\begin{subfigure}{0.48\textwidth}
  \centering{\includegraphics[height = 4.5cm,width=4.5cm]{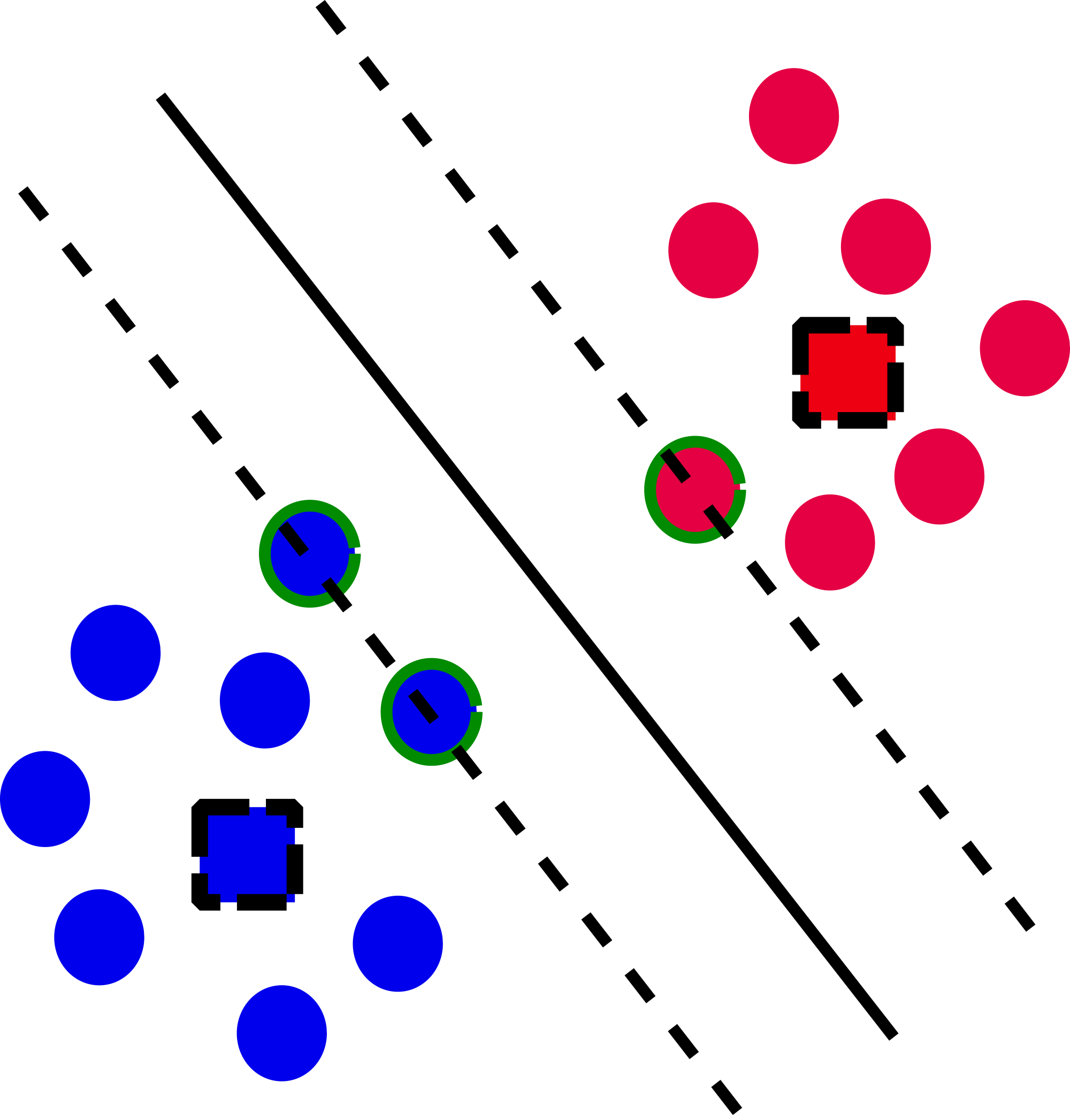}}
  \caption{Linear Support Vector Machine }
 \label{fig:svm}
\end{subfigure}

\begin{subfigure}{0.48\textwidth}
  \centering{\includegraphics[height = 4.5cm,width=4.5cm]{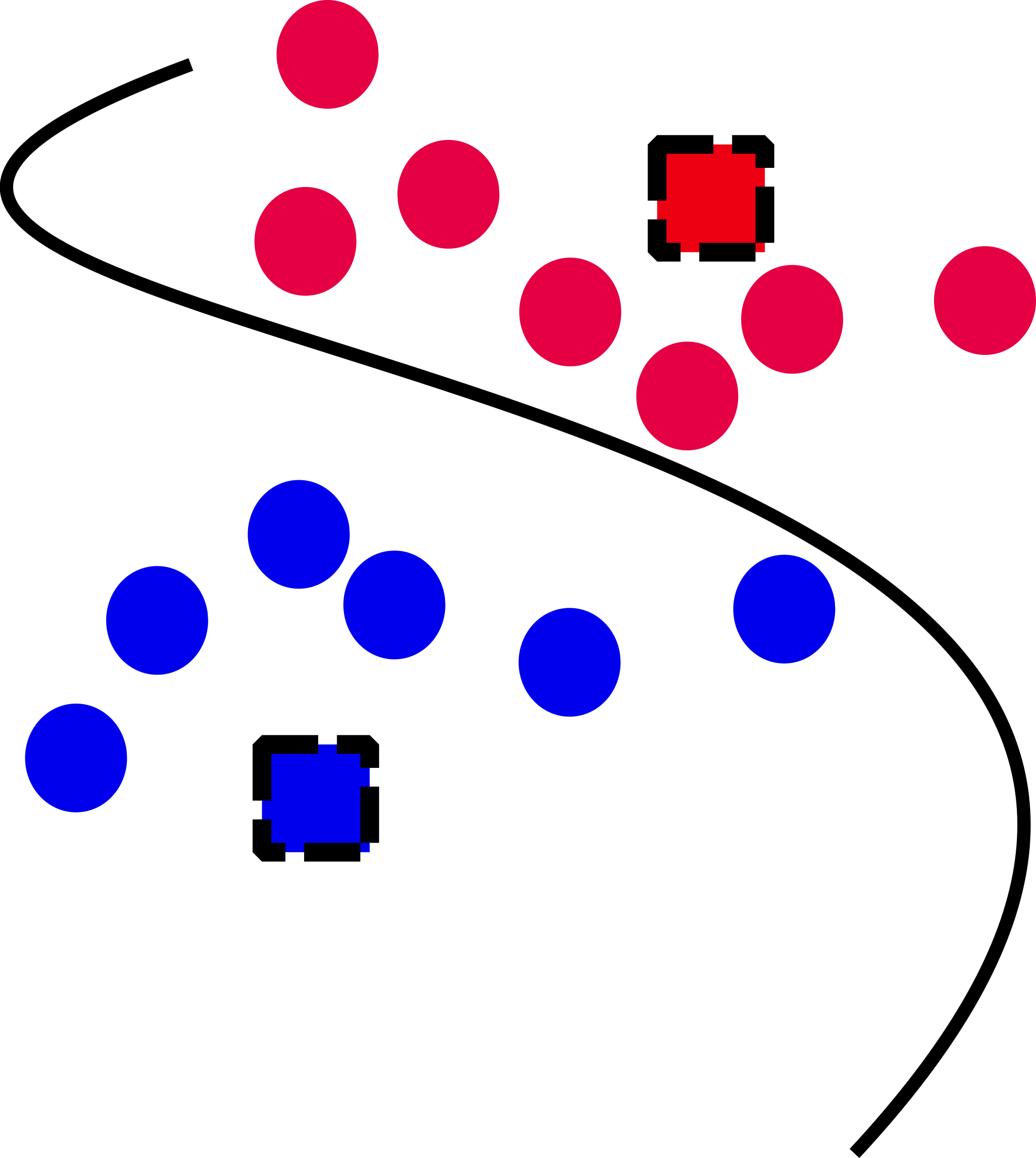}}
  \caption{Non-linear SVM classifier with soft margin }
  \label{fig:kerneltrick}
  \end{subfigure}
    \caption{\ac{SVM} for proximity based fingerprinting}
\end{figure}

First a training dataset, linearly separable, is given consisting of data points $\mbf x_i\in \mbb R^p$ with labels $y_i\in\{-1,+1\}$. The idea of \ac{SVM} is to use a hyper-plane $f(\mbf x)=\mathbf{a}^T\mbf{x}+b$ to separate two classes so that each class lies on one side of the hyper-plane:
\begin{equation}
    y_i (\mathbf{a}^T\mbf{x}_i+b)\geq \gamma >0,
\end{equation}
for some $\gamma>0$. The optimal hyper-plane would be the one which maximum margin between two classes. It can be seen that the following optimization problem provides a solution for $\mbf a$ and $b$ \cite{steinwart_support_2008}:
\begin{equation}
    \arg\underset{\mbf a,b}\min\frac{\|\mathbf{a}\|^2}{2}\quad \mrm{s.t.} \quad y_i(\mathbf{a}^T\mbf{x}_i+b)-1\geq 0.
\end{equation}
One can equally solve the dual problem by considering the Lagrangian, yielding the following problem:
\begin{equation}
\begin{aligned}
        \arg\underset{\boldsymbol\lambda}{\max}\sum_{i=1}^n\lambda_i-\frac{1}{2}\sum_{i=1}^n\sum_{j=1}^n \lambda_i\lambda_j y_i y_j\mathbf{x}_i^T\mbf{x}_j 
\\\text{s.t.} \sum_{i=1}^n\lambda_iy_i=0\quad \text{and}\quad 0\leq \lambda_i.
\label{fig:SVMdual}
\end{aligned}
\end{equation}
Note that the dimension of search space for the dual problem scales with the size of training set but for the primal problem  with the dimension of training points. Therefore, although both problems can be considered as quadratic optimization problem and can be easily solved by quadratic programming algorithms, the choice of which problem to solve depends on then number of training points and their dimension. After solving the dual problem, $\mbf a$ can be obtained as $\mathbf{a}=\sum_{i=1}^{n}\lambda_iy_i\mbf x_i$. The support vectors with are those with $\lambda_i>0$ and they solely determine $\mbf a$ and $b$. For a support vector $\mbf x_k$, $b$ is obtained as $b=y_k-\mbf a^T\mbf x_k$. Th support vectors are shown in Fig.\ref{fig:svm}.

In order to deal with non-separable data, a penalty term is added which tries to minimize the number of points inside the margin. If the margin violation of each point is denoted by $\xi_i$, the goal is to minimize the $\ell_0$-norm of $\boldsymbol\xi=(\xi_1,\dots,\xi_n)^T$ for this purpose. To have a convex formulation $\ell_1$-norm is used instead  which is well known to provide sparsity. Therefore the following optimization problem is solved:
\begin{align*}
\arg\underset{\mbf a,b}\min&\frac{{\|\mbf a\|}^2}{2}-C\sum_i \xi_i \\
\quad & \text{s.t.} \quad y_i(\mathbf{a}^T\mbf x_i+b))\geq 1-\xi_i \text{ and } \xi_i\geq 0.
\label{softmargin}
\end{align*}
where $C$ is a parameter which should be correctly chosen. The equivalent dual problem is given by:
\begin{equation}
\begin{aligned}
& \arg\underset{\boldsymbol\lambda}{\max}
& & \sum_{i=1}^{n} \lambda_i -\frac{1}{2} \sum_{i=1}^n\sum_{j=1}^n  \lambda_i \lambda_j y_i y_j\mathbf x_i^T\mathbf x_j \\
& 
& &\text{s.t}\quad \sum_{i=1}^{n} \lambda_i y_i = 0\text{ and } 0 \leq \lambda_i \leq C.
\end{aligned}
\label{eq:Dual-nonsepSVM}
\end{equation}
Note that the search space has the dimension $p+n+1$  for the primal problem while the dimension of  the dual problem remains unchanged for both separable and non-separable case, equal  to $n$ hence making it more efficient to solve for non-separable case. 

As another advantage of the dual problem, it can be easily extended to a non-linear SVM classifier which is achieved by using kernel trick. In this case the inner product $\mathbf{x}_i^T\mbf x_j$ as in \eqref{eq:Dual-nonsepSVM} is replaced by a kernel function $K(\mathbf{x}_i,\mbf x_j)$. The kernel function represents an inner product of the transformations of these vectors in a feature space, usually higher dimension and possibly infinite dimensional. This transformation is only done implicitly and in many cases the transformation function is not explicitly known. Since the dual problem only depends on the inner product, it is sufficient to know the inner product in the feature space as a function of training points and kernel functions provide this information. By transformation into higher dimensional space, a linearly non-separable data in the ambient space might become linearly separable in the feature space. An example is given in \eqref{fig:kerneltrick} where a linearly non-separable data is classified using kernel tricks. Some examples of kernel functions are polynomial kernel, \ac{RBF} kernel and hyperbolic tangent kernel. The kernel functions contain design parameters too. For instance, \ac{RBF} kernel is given by $K(\mbf x,\mbf y)=\exp(-\gamma\|\mbf x-\mbf y\|^2)$ where $\gamma$ is a design parameter. In its most general form, \acp{SVM} only require to tune few parameters namely the constant $C$ in \eqref{eq:Dual-nonsepSVM} and parameters of kernel function. In its most general form, \ac{SVM} classifier is constructed by solving the following problem
\begin{equation}
\begin{aligned}
& \arg\underset{\boldsymbol\lambda}{\max}
& & \sum_{i=1}^{n} \lambda_i -\frac{1}{2} \sum_{i=1}^n\sum_{j=1}^n  \lambda_i \lambda_j y_i y_j K(\mathbf x_i,\mathbf x_j) \\
& 
& &\text{s.t}\quad \sum_{i=1}^{n} \lambda_i y_i = 0\text{ and } 0 \leq \lambda_i \leq C.
\end{aligned}
\label{eq:generalSVM}
\end{equation}
Note that since we are working in the feature space, $\mbf a$ is really $\sum_{i=1}^n \lambda_i\Psi(\mathbf x_i)$ where $\Psi$ is the feature mapping. However the mapping is only implicitly known and hence $\mbf a$ cannot be known although the classifier itself can be constructed. First $b$ is obtained as
\begin{equation}
 b=y_k-\sum_{i=1}^n\lambda_iK(\mbf x_i,\mbf x_k).
\label{eq:bSVM}
\end{equation}
Consequently the classifier for a new observation $\mathbf x$ is constructed as
\begin{equation}
 \sum_{i=1}^n\lambda_iK(\mbf x_i,\mbf x)+b \gl^{y=1}_{y=-1} 1.
 \label{eq:genSVMclassifier}
\end{equation}

\subsection{SVM in Fingerprinting Algorithm}
In the context of fingerprinting algorithm, the fingerprint of each training point is a sequence of equal number of measurements from different anchors. Therefore for $A$ anchors and $M$ measurements per each, the training point fingerprint $\mathbf X$ is a vector of dimension  $A\times M$. 
To apply \ac{SVM} in fingerprinting algorithm, first the training points are chosen. The Voronoi diagram corresponding to the grid divides the location space into different regions where each region can be seen as a class and all points in one region are seen as one class with the training point as the label. The aim is to classify the test point to one of those partitions using \ac{SVM}. 

Note that without proximity data, only a single observation is available for a given class. If so, \ac{SVM} would be exactly the same as Euclidean based distance localization. It is also possible to obtain multiple observations using data augmentation which will be discussed later. Casting the localization problem as classification makes possible to utilize the so called imprecise measurements. In that case, it would be enough to know the region in which observations are collected, i.e., the label of measurements. Unlike training points on the grid, no precise location information is needed. Interestingly this improves the localization performance. Therefore for each training point $\mathbf v$, there are precise measurements and proximity measurements corresponding to its Voronoi region. One needs particularly to solve a multi-class classification problem with number of classes equal to the cardinality of the training grid $\Lambda$. Therefore multiple binary \ac{SVM} classifiers have to be trained \cite{hsu2002comparison} and the final decision is made according to  \textit{one vs. one} or \textit{one vs. rest} strategy. 

For $|\Lambda|=k$ classes, in one vs. one approach, $\binom{k}2$ binary classifiers are constructed and the class with highest number of decisions is chosen. The steps are as follows.
\begin{enumerate}
\item Consider all fingerprints of two training points $\mathbf v_k$ and $\mbf v_l$ in $\Lambda$. The fingerprint  $\mathbf X^{(kl)}$ is labeled with $y=1$ if it corresponds to $\mathbf v_k$ and $y=-1$ if it corresponds to $\mathbf v_l$. 
 \item Solve the optimization problem \eqref{eq:generalSVM} to find $b_{(kl)}$ and the following classifier for a fingerprint $\mbf X$:
\begin{equation}
 \sum_{i=1}^{n^{(kl)}}\lambda_i^{(kl)}K(\mbf X^{(kl)}_i,\mbf X)+b_{(kl)} \gl^{y=1}_{y=-1} 1.
 \label{eq:genSVMclassifierFP}
\end{equation}
\item For a fingerprint $\mbf X$, the output of the above classifier is $\mbf v_{(kl)}$ which is equal to either $\mbf v_k$ or $\mbf v_l$.
\item Repeat the above steps for each pair of training points. 
\item The final decision is given by
$$
\mathbf v_{\mbf X}=\arg\max_{\mathbf v\in\Lambda}\sum_{i,j}\mathbf 1(\mbf v_{(ij)}=\mbf v).
$$
\end{enumerate}
In one vs. rest approach only $k$ classifiers are trained and for each classifier one class is tested against all other classes put into a single one. The final output is also based on majority decision. Similar to traditional fingerprinting approach, post-processing like \ac{kNN} can be also used in \ac{SVM} approach. Instead of choosing the best class, multiple top classes can be chosen. The final location would be the average of corresponding training points location. 

\section{Deep Learning of Localization Function and Fingerprint Construction}
\label{sec:DeepLearning}
\begin{figure}[h]
  \centering{\includegraphics[width=0.5\textwidth]{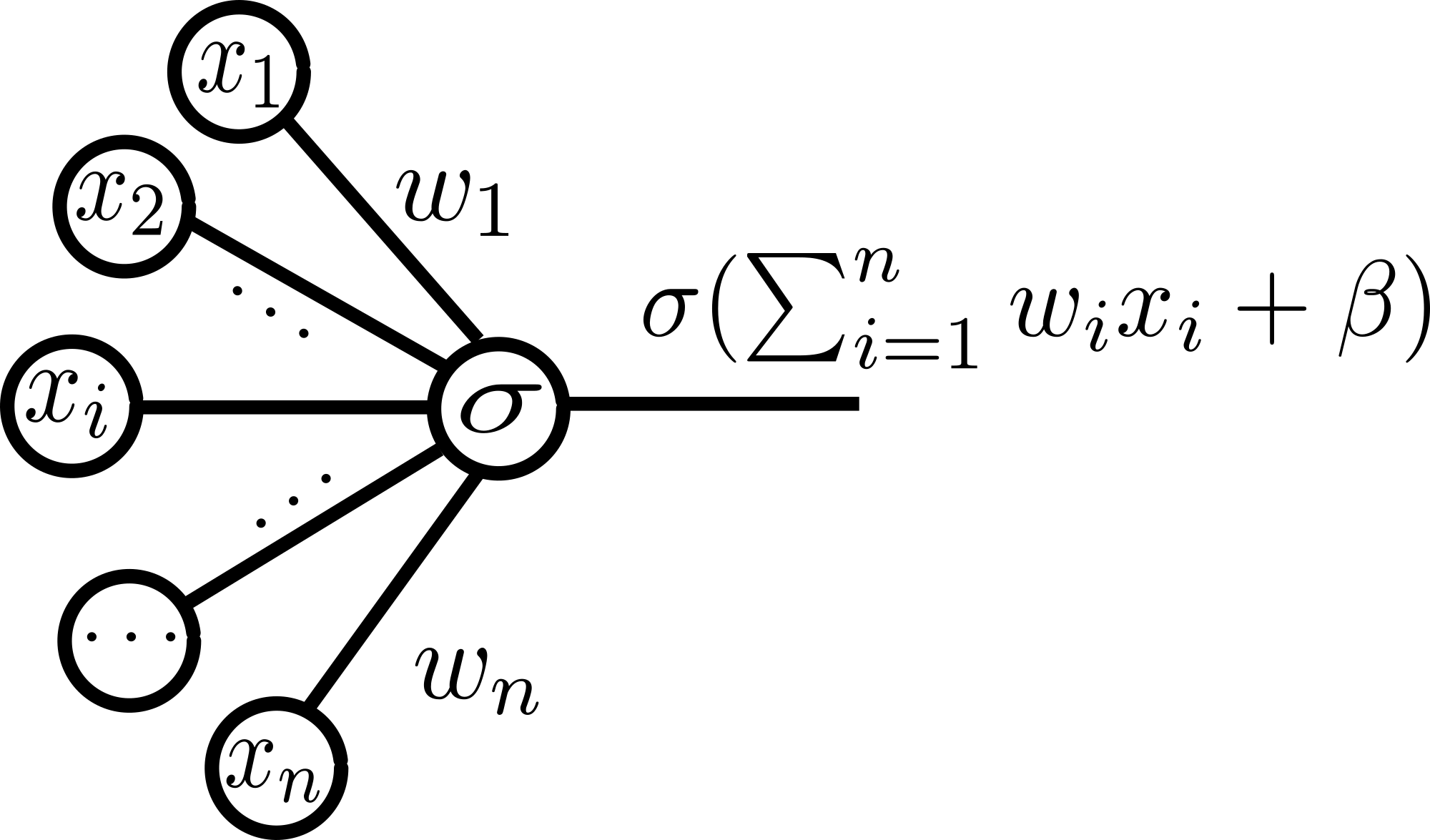}}
  \caption{An artificial neuron }
    \label{fig:nnperceptron}
\end{figure}
Deep learning architectures emerged as the prime candidate for complex learning problems with excellent performance during recent years. In span of few years these architectures outperformed conventional machine learning algorithms in tasks such as pattern recognition.  Neural networks consist of multiple units called neurons connected to each other where each neuron computes a function of its input value, mostly a non-linear function. The input to each neuron is the linear combination of the output of some other neurons. The basic building block is an artificial neuron, Fig. \ref{fig:nnperceptron}. The neuron computes the function $\sigma(\mathbf w^T\mbf x+\beta)$ where the function $\sigma(.)$ is called the activation function. The main question about the capabilities of this architecture in learning different functions. A disappointing answer is that a single neuron might not be able to learn a function like XOR function \cite{minsky_perceptrons_1972}. This looks however more promising when one focuses on multi-layer case. 
A neural network with a single hidden layer is capable of learning any  function on the set of continuous functions over $n$-dimensional cube $I_n=[0,1]^n$ denoted by $C(I_n)$   \cite{cybenko_approximation_1989,hornik_approximation_1991}. More precisely, the classic result states that for a class of functions $\sigma(.)$ and for any function $f\in C(I_n)$ and $\epsilon>0$, there is a finite sum of the form $G(\mbf x)=\sum_{i=1}^n\alpha_i\sigma(\mathbf w_i^T\mbf x+\beta_i)$ such that $|G(\mbf x)-f(\mbf x)|\leq \epsilon$ for all $\mbf x\in I_n$. The class of activation functions $\sigma(.)$ contains continuous discriminatory functions. This result has been extended recently to a class of unbounded activation functions yielding the approximation of functions $L^1(\mathbb R^n)$ in pointwise convergence sense \cite{sonoda_neural_2017}.

Although promising, it is a very challenging to find the correct number of neurons and weights for a specific function. Usually the parameters are found based on a dataset containing input-output samples of the desired function. The process of learning the weights to be able to perform certain tasks is called training phase.
The training is done usually through a procedure called back propagation where the weights are adjusted iteratively to minimize the output error for a training set. The main challenge in training neural networks is that the output error as a function of weights contains many local minima and saddle points, making it extremely difficult to find the global optimum.
Although theoretically a single hidden layer would suffice for approximation, it has been recently shown that multiple hidden layers can perform tasks that require exponentially bigger number of neurons if it is done in a shallow architecture \cite{raghu_expressive_2017}. Deep learning refers to an architecture in which the neurons are organized in many consecutive layers. The training of these architectures was the main obstacle for their development, which has been circumvented during recent years. Deep learnings now are used for performing supervised and unsupervised learning tasks as well as dimensionality reduction and feature extraction tasks.

In this section, deep neural networks are used for two main task, first to extract essential features of \ac{RSSI} measurements and second  to perform regression task for indoor localization problem. The goal of regression analysis is to learn the localization function $\Phi$, approximated using the neural network for mapping fingerprints  to the estimated location of this point. The training set is therefore the pairs of training points and their fingerprints. Compared to \ac{kNN} approach, this approach combines pattern matching and post-processing and provides the location in one shot. On the other hand, the received \ac{RSSI} values are not averaged and directly used as the input to the algorithm. This might prevent the possible information loss in averaging.
In this work, we focus on design issues including the influence of different hyperparameters, avoiding overfitting and training algorithms. The hyperparameters in deep learning consists of number of layers, number of neurons in each layer, the choice of non-linearity parameters, learning rate, etc. 
The choice of hyperparameter is an important problem in deep learning and currently there is almost no unified theory for choosing those parameters. However there are many guidelines derived from vast experimental researches  such as in \cite{bengio2012practical}. For simplicity, not all the hyperparameters are discussed in this paper. More focus is put into those hyperparameters with seemingly most important effect.

\subsection{Fingerprint Construction using Autoencoders}

\begin{figure}[h]
  \centerline{\includegraphics[height = 4.1cm,width=8cm]{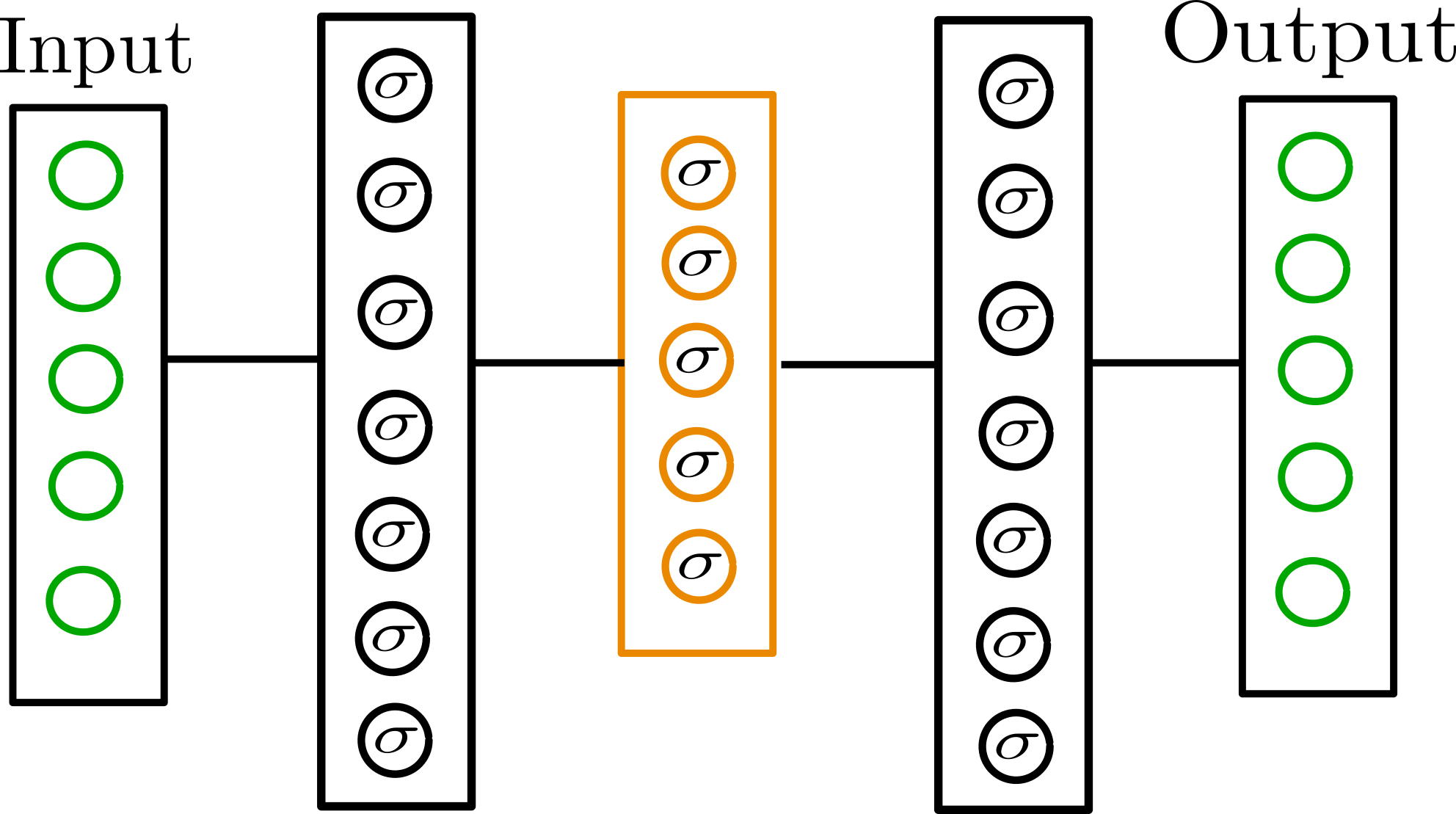}}
  \caption{Autoencoders }
    \label{fig:AE}
\end{figure}
Autoencoders, also known as auto-association, was historically construed as a model for memory. Mainly an unsupervised learning architecture, it aims at finding a representation of data in a space of lower dimension. An autoencoder, Fig. \ref{fig:AE} is a multi-layer neural network which attempts at constructing the very same input by mapping it into hidden layers and ultimately finding a low dimensional representation of data. In that sense they are used for feature extraction and dimensionality reduction. Interestingly, an autoencoder with merely linear layers performs nothing but \ac{PCA} for dimensionality reduction \cite{linsker_self_organization_1988}. Although proposed decades ago, autoencoders emerged again as they were used for layer-wise pre-training of deep neural networks \cite{bengio2007greedy}. Its variants such as stacking autoencoders and denoising autoencoder were proposed 
 \cite{vincent2008extracting,vincent2010stacked} where the latter is trained on corrupted inputs and therefore can be used as denoising tool too. 
Typically, the encoder gradually maps the input to a lower dimensional space through multiple hidden layers of decreasing size. The decoder symmetrically consists of layer of increasing size. Training of autoencoders follow a similar procedure to conventional neural networks. The low dimensional representation of the data is obtained by applying the encoder to test data and then fed into other classification algorithms. 

In a similar fashion, autoencoders can be used for feature extraction from \ac{RSSI} values. As it was discussed before, generally the \ac{RSSI} values from each \ac{AP} is averaged out and therefore all measurements are stored in a vector of dimension equal to the number of \acp{AP}. The question is whether this naive approach is sufficient for either good compression or effective feature extraction. The measured \ac{RSSI} values are used as the input to the autoencoder and through an encoder of single hidden layer its dimension is reduced. Three main issues should be considered here, first data preparation, second hyperparameter choice such as the dimension of hidden layer and the number of hidden layer neurons and finally the training of hidden layer weights. These steps are common to neural networks and will be discussed extensively in later sections.

\subsection{Fingerprinting-Location Regression using Deep Learning}
\begin{figure}[h]
  \centerline{\includegraphics[height = 4.1cm,width=8cm]{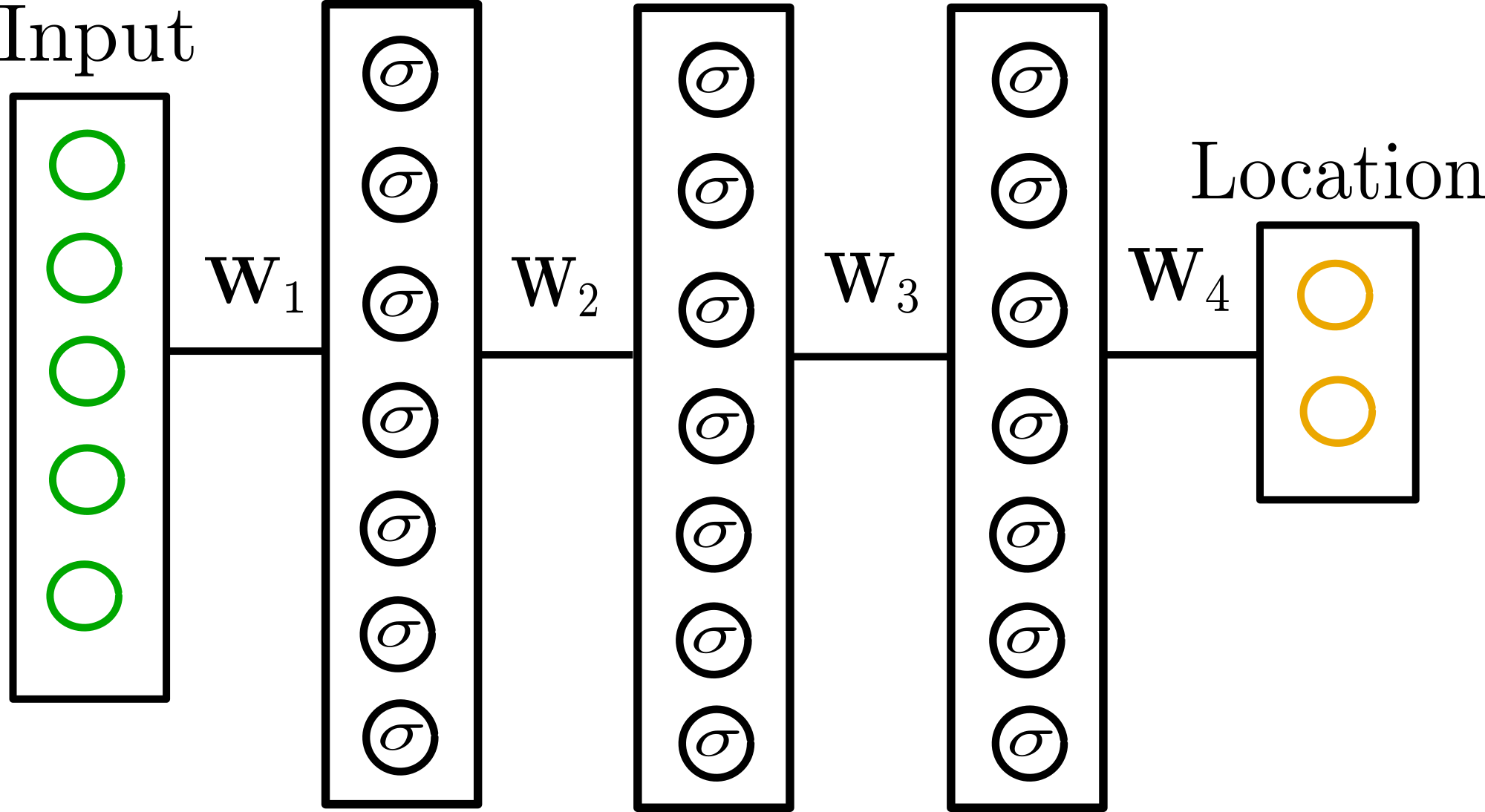}}
  \caption{\scriptsize Neural network configuration }
    \label{fig:nndesign}
\end{figure}

With complete and precise knowledge of propagation environment as well as transmitters property, it is possible to find the received powers at different locations using elegant but rather complex derivations. However it is difficult to have actual and complete information in a complex indoor environment. On the other hand, the localization requires solving an inverse problem which maps received powers to the location. This is even more difficult considering the model complexity.  An alternative approach is to approximate this function, i.e., the localization function $\Phi$, using many samples.

In this work, deep neural networks are trained to approximate the localization function $\Phi$, Fig \ref{fig:nndesign}. The input is the raw \ac{RSSI} values directly obtained from different \acp{AP}. The output is the estimated coordinates of the test point. The size of the output layer is fixed therefore either two or three dimensional space. The input size however cannot be chosen a priori fixed. The number of \ac{RSSI} measurements from an anchor can differ from point to point particularly when their acquisition is dependent upon correct reception of the packet. Therefore the measurements can have different size at each point while the input size of the neural networks should be fixed. This problem is addressed in the next section where a solution is proposed for adjusting the number of measurements to the input size.

\section{Designing Deep Neural Networks}
\label{sec:design}

Deep neural networks contain huge number of parameters and hyperparameters. Each one of them is chosen differently from one problem to another. Although the neural network design still lacks a unified theory for parameter selection, there are many guidelines and insights obtained throughout the years to address common problems arising in applications. In what follows, we review and employ some of these insights in context of indoor localization.

\subsection{Resampling and Normalization}
\ac{RSSI} measurements are used as the input to the neural networks. There are two main problems about directly using the measurements for the input. The first problem is a practical one. The number of measurements from different \acp{AP} differ at each training point. This is a recurring problem when the acquisition of \ac{RSSI} values is dependent on the correct reception of the packet. Therefore when \ac{SNR} of the received packet is not suitable for correct reception, one receives a few or even no packet to have \ac{RSSI}. In this case, receiving more packets amounts to further attempts for correct reception and increased latency. To address this problem, we propose a resampling method based on bagging \cite{leo_breiman_bagging_1996}. Bagging method aggregates the output of multiple learning models, each on fit to a different dataset randomly sampled from the original one. The idea of constructing multiple randomly sampled versions of a given dataset is introduced for precision analysis of models which includes techniques like bootstraping  \cite{leo_breiman_bias_1996,zhou_ensemble_2012}. Aggregating multiple models can lead to a better model built out of many simpler one while using random sampling lead to lower variance 
by diminishing the effect of accidental regularities. 

As it will be shown later, multiple model aggregation is embedded into deep neural networks by proper regularization. In that way, a deep neural network can be seen as an ensemble of learners working in parallel joint with an aggregation at the end. However in this work, the random sampling is particularly implemented but from measurements of each \ac{AP}. Suppose that at a training point $\mbf v$,  $N_k(\mbf v)$ measurements are available from the anchor $k$. $N$ samples are randomly taken with replacement from these $N_k(\mbf v)$ measurements and put in the resampled fingerprinting database. This process is repeated for each training points and each anchor. When no measurement is available from an anchor, a default value is 
used which corresponds to the minimum \ac{RSSI} value. This procedure leads to a database containing equal number of measurements from each anchor.  

The second problem is about the scaling of data. \ac{RSSI} measurements, measured in Watt, belong to positive real numbers and can be very large if very close to the anchors. These large values lead to gradient vanishing problem for non-linearities like sigmoid function. In general, without proper normalization of error, weights or the input, these values start from values belonging to different order of magnitudes and the back-propagation algorithm might not even converge. One solution is normalize all \ac{RSSI} values between zero and one. Suppose that $X_{\min}$ $X_{\max}$ are respectively the minimum and maximum measured \ac{RSSI} value. Then the measurement $X$ is normalized using $\frac{X-X_{\min}}{X_{\max}-X_{\min}}$. 
The idea of normalization appears also in image applications. The images might have different contrasts and hence creating variations not essential to the task. The \ac{GCN} is used to avoid varying contrasts by subtracting the mean and rescaling to get equal standard deviation per pixel \cite[Chapter 12]{goodfellow_deep_2017}. Note that a similar normalization cannot be used here since the variance and the mean value contains exactly those information vital for localization.

\subsection{Activation Function}
For a long time, logistic sigmoid function $\frac{1}{1+e^{-x}}$ has been the common choice of activation function. The problem is that for inputs of large absolute value, the gradient function is very small and therefore the gradient do not affect the weights in back-propagation. This problem is called vanishing Gradient problem. Recently, the favorite choice for non-linearity is \ac{ReLU} function, defined as $f(x)=\max\{0,x\}$. It does not suffer from saturation and converges faster than sigmoid to an acceptable minimum \cite{jarrett_what_2009,glorot2011deep,krizhevsky2012imagenet}.

\subsection{Number of Hidden Layers and Neurons}
There is currently no systematic method to choose the number of hidden layers and neurons. However, the common belief is that higher number of hidden layers are capable of approximating more complex functions \cite{bengio2012practical}. There are some theoretical works supporting this claim \cite{bengio_learning_2009,raghu_expressive_2017}. 
These works show that certain class of functions, or operations, can be implemented using deep networks while a shallow network would require an exponentially higher number of neurons to do the same. Therefore utilizing more hidden layers expands the expressiveness and hence, one can approximate more complex functions. However the difficulty of training increases as the number of parameters increases. In general, the complexity of the model, in this case the neural network, should match to the complexity of the data. A priori function about the complexity of the operation might hint to the choice of model. For instance, the binary classification for a dataset which is linearly separable can be done using a single layer Perceptron without any hidden layer. However the information about the complexity is not enough. If the training data is not big enough to capture the complexity of underlying structure, the choice of complex model for learning would lead equally to overfitting problem.  Same rules of thumb apply to the choice of number of neurons. In general more training data encourages more number of neurons. In general the number of neurons in each hidden layer should not significantly exceed the input size, to avoid overfitting,  and should not be significantly smaller than the output size, to avoid underfitting. 

\subsection{Weight Initialization}
As it will be discussed later in the next section, the training of neural networks consists of minimizing the error function as a function of all weights. This function contains many local optimum and saddle points and it is notoriously difficult to find a good local optimum let alone finding the global optimum. Since the training is done using gradient descent-based update of weights, the initial value of weights can affect significantly the success of training algorithms.
A breakthrough in deep learning research came with the idea of layer-wise pre-training of the network weights \cite{hinton2006fast}. The idea is to pre-train the weights of neural networks in order to put them in a good starting point in error space and then fine-tuning the whole network following forward, backward propagation update procedure. Recent works have shown that a straightforward initialization of weights suffices for satisfactory training \cite{glorot2010understanding,he2015delving} if it is chosen with attention to the choice of non-linearities and proper possible input values \cite{mishkin_all_2016}.  In this work, the weight matrix of $i$-th layer  $\mbf W^{(i)}$  is a random matrix with i.i.d. entries following Gaussian distribution $\mathcal{N}\left(0,\sqrt{\dfrac{2}{n^{(i)}}}\right)$ where  $n^{(i)}$ is the number of neurons in $i$-th layer \cite{he2015delving}.  

\subsection{Gradient-based training of weights}
Neural networks aim at approximating functions using consecutive application of linear transformations and non-linearities. An important design choice in this context concerns linear transformations simply called weights here. Once the weights are initialized, the weights should be chosen to minimize the error function for the training set defined as a function of weights $E(\mbf W)$. If the explicit characterization of the approximation error as a function of weights were at hand, one could use global optimization techniques to minimize the error and find the optimal weights. However in many learning applications, the desired function is unknown. The function is characterized using its input-output instances provided by the training set. Therefore one can only measure how well a neural network is capable of producing similar input-output instances. An error function for the network is specified for that purpose. The error function is minimized for the instances in the training set using an iterative procedure. The weights are then updated in each iteration to minimize the error mostly using Gradient-based update rules. This means that at each iteration, the gradient of the error function is calculated for all weights and  used to update the weights. 
 
Since the error function should be minimized for all the training examples, one might be tempted to minimize the error function for each training instance at each iteration and repeat this multiple times over the whole training set. This is called \ac{SGD}. However the dependence of \ac{SGD} on one single training example does not necessarily guarantee that the gradient of the single example aligns necessarily with the direction that minimizes the error for all samples. Consequently, the gradient updates behave in an oscillatory fashion and therefore lead to significant increase in the training time. Another approach is to use \ac{BGD}, where the error function is the sum of errors of all training instances. There are some problems with this approach. First, when the number of training instances are big the training takes a very long time. The second problem appears particularly when \ac{BGD} is used for training autoencoders. The mean squared error with \ac{BGD} is equivalent up to a scaling factor to the average error. The error is minimized if the autoencoder learns to produce the empirical average of the data which is insufficient as representation of the whole data.

To avoid the gradient oscillation of \ac{SGD} and improve \ac{BGD}, the training set is divided into so called \textit{mini-batches} each one containing $B$ training points. At each step, the error is minimized for a single mini-batch and weights are updated proportional to  their contribution to the mini-batch error. The process is called \ac{MBGD}. Note that  \ac{SGD} corresponds to the case $B=1$ and \ac{BGD} to the case $B$ equal to the whole training data-set. Typical mini-batch sizes are 32, 64, 128 \cite{lecun2012efficient}. \ac{MBGD} has two main advantages. The distribution of a mini-batch of data fits the distribution of the whole data better than a single example and it is converging faster than \ac{BGD}. The trade-off of above algorithms can be summarized between accuracy and time required for one iteration which is discussed in details in \cite{lecun2012efficient}. 

In all the above approaches, the training is repeated over the training set for multiple times, called epochs. At iteration $n$, the information about the effect of each weight on the total error is encapsulated in the partial derivatives $ \frac{\partial E(\mbf W^{(n-1)}) }{\partial \mbf W^{(n-1)}}$. These derivatives determine the gradient direction. The common method would be to update the weights by descending on the reverse gradient direction, i.e.,  by subtracting $\alpha \frac{\partial E(\mbf W^{(n-1)}) }{\partial \mbf W^{(n-1)}}$ from the weights. The term $\alpha$ is called the learning rate and it is an important design choice.

However this approach, conventionally used in the neural networks for many years, faces huge difficulties. The main difficulty is related to the particular shape of the error function. The function is a high-dimensional non-convex function and contains many local minima as well as  many saddle points. A training algorithm can easily get stuck in a local minima or saddle point if the parameters are not chosen properly. The first parameter is the initialization of weights which constitutes the initial value for the Gradient-based algorithms. An initialization sufficiently close to the global minima or a good local minima plays a vital role in success of the training algorithm. 

In \cite{polyak1964some}, a term called momentum inspired from physics was introduced to account for the memory of previous gradient updates and alleviate the oscillation problem. In this case, the update is not only depending on the current mini-batch but also the last mini-batch. A momentum parameter $\rho$ needs to be chosen to control how much the last update is taken into account. Typical decaying parameter equals 0.9 or 0.99. At iteration $n$, the weights are updated by $\Delta \mbf W^{(n)}$ which is determined by 
\begin{equation}
\Delta \mbf W^{(n)} = \rho \Delta \mbf W^{(n-1)}- \alpha \frac{\partial E(\mbf W^{(n-1)}) }{\partial \mbf W^{(n-1)}}.
\end{equation}
The update is given by $\mbf W^{(n)} = \mbf W^{(n-1)}+\Delta \mbf W^{(n)}$.  

The learning rate $\alpha$ and the momentum control parameter $\rho$ should be carefully chosen. The learning rate controls the step size of gradient descent. A very large learning rate can cause the updates oscillating around the minimum or even diverge. On the other hand a very small learning rate will make the updates very slow and possibly getting stuck in a local minima. There are many methods for better tuning of the learning rate and we will discuss some of them here. Although the same learning rate can be adopted for updating all weights,  AdaGrad \cite{duchi2011adaptive} was designed to enable different weights to have different learning rates.  A uniform initial learning rate $\alpha$ is chosen for all weights at the beginning. The learning rate at each iteration is divided by square root of sum of squares of gradients for previous iterations. 
Therefore the weight updates decay in time inversely proportional with the Gradient of previous steps. A problem of AdaGrad is that the decaying factor is accumulative and hence monotonically increasing in time. Therefore the learning rates are monotonically decreasing and at the end all learning rates will go down to zero. RMSProp \cite{tieleman2012lecture} was introduced in 2012 to prevent such effects. Instead of simply adding up the squares of all the gradients, the new decaying factor for the learning rate is obtained by multiplying the decaying factor of previous steps by a parameter $\gamma<1$. Therefore the effect of recent iterations will be more significant than the early iterations. The rate $\gamma$ is normally chosen to be 0.9. Adam \cite{kingma2014adam} was designed in 2014 to combine both benefits of momentum and adaptive learning rate. Adam is easy to implement and computationally efficient. Similar to momentum and RMSProp, decaying parameter $\gamma$ and momentum parameter $\rho$ need to be chosen. The authors suggest $\rho$= 0.9 and $\gamma$ = 0.999 as the default values. In the first few iterations, a bias correction has to be applied to prevent the update from going wrong when the momentum and the decaying factors are initialized at zero.
Adam has been proven to be a very powerful variant of gradient descent due to the fact that the algorithm is not that sensitive to the initial learning rate.
In this work Adam is chosen as our default optimization method.


\subsection{Regularization}

A central problem in machine learning is overfitting. The typical sign of overfitting is when the learning algorithm offers a very good performance on the training data but it performs badly on the test data. This is because the learning algorithm utilizes a  model which is more complex and therefore learns those features of particular training set which are non-essential to the task.
Often this is due to the high number of free parameters in the model.  The problem is particularly grave for neural networks since even a simple multilayer neural network contains way more parameters that the input dimension. Regularization is an important technique in statistics particularly in the context of inverse problems used for adding more constraints on the desired solution. By limiting the variation of free parameters, it can be used to prevent overfitting.

\subsubsection{Early Stopping}
The gradient descent algorithms make sure that the total training error is decreasing with the number of iterations. But a small training error is not equivalent to a small test error. Too many iterations of gradient descent may even damage the performance of the system instead of improving it. Early stopping is an inexpensive scheme to prevent the neural network from being overtrained. The idea is to stop the training when the validation error is not decreasing significantly with iterations. 

When implementing the early stopping, a parameter called patience $P$ has to be chosen manually which is the number of iterations to update after seeing a minimum. As training proceeds and a new minimum in validation error with its weights setting $\mathbf{W}^{(T)}$ are observed at the $T^{th}$ iteration, $\mathbf{W}^{(T)}$ will be saved in memory as the best candidate model. The update will still continue till the $(T+P)^{th}$ iteration. If a even lower validation error is observed during $P$ times iterations, then the system model will be updated and another $P$ times iterations will be proceeded. This procedure will stop until no better model found during $P$ times iterations. The patience is set to avoid immediate stop when there is an oscillation of the error.
 The use of early stopping makes the number of training iterations a hyperparameter that can be easily optimized.

\subsubsection{$\ell_1$ and $\ell_2$ Regularization}
In neural networks, one way to restrict the free choice of all parameters is to introduce constraints on them. This approach is based on using either $\ell_1$ or $\ell_2$-regularization.
In this case, a penalty term $\displaystyle\lambda \sum_{i=1}^L \|\mbf W_i\|_p^p$, $p=1$ or $2$, is added to the error function of the neural network to be minimized. The penalty term restricts the norm of weights to be small for $p=2$ and promotes sparsity of the weights for $p=1$. $\lambda$ is a parameter to control the degree of penalty. 
The optimization problem for $\ell_2$ is changed to:
\begin{equation}
\mathop{\min}\limits_{\mathbf{W}}(E(\mathbf{t}, \mathbf{y},\mathbf{W})+\frac{\lambda}{2}\|\mathbf{W}\|_2^2),
\label{regularizedl2 objective function}
\end{equation}
with the corresponding gradient:
\begin{equation}
\lambda \mathbf{W} +\nabla_\mathbf{W} E(\mathbf{t}, \mathbf{y},\mathbf{W}).
\end{equation}
Thus the update of gradient descent step is given by:
\begin{equation}
\mathbf{W}^{(n)} = (1-\alpha\lambda)\mathbf{W}^{(n-1)} - \alpha \nabla_{\mathbf{W}^{(n-1)}} E(\mathbf{t}, \mathbf{y},\mathbf{W}^{(n-1)}).
\end{equation}
It can be seen that the $\ell_2$ regularization shrinks the weight vector by a constant factor $(1-\alpha\lambda)$ for each step before usual gradient update.
The optimization problem for $\ell_1$ regularization is given by:
\begin{equation}
\mathop{\min}\limits_{\mathbf{W}}(E(\mathbf{t}, \mathbf{y},\mathbf{W})+\lambda\left \| \mathbf{W} \right \|_1),
\label{regularizedl1 objective function}
\end{equation}
with the corresponding gradient update:
\begin{equation}
\lambda \mathrm{sign}(\mathbf{W})  +\nabla_\mathbf{W} E(\mathbf{t}, \mathbf{y},\mathbf{W}),
\end{equation}
where $\mrm{sign}(\mathbf{W})$ is the sign of $\mathbf{W}$ applied element-wise. The update of one step for $\ell_1$ regularization is:
\begin{equation}
\mathbf{W}^{(n)} = \mathbf{W}^{(n-1)}-\alpha\lambda \mathrm{sign}(\mathbf{W}^{(n-1)})  - \alpha \nabla_{\mathbf{W}^{(n-1)}} E(\mathbf{t}, \mathbf{y},\mathbf{W}^{(n-1)}).
\end{equation}


\subsubsection{Dropout}

Another regularization technique is called  dropout which prevents overfitting by randomly dropping some hidden units during weight updates for each iteration \cite{srivastava2014dropout}. In this way, the intermediate representations of the input is not dependent on only few neurons and the network tends to learn a distributed representation of the data.  The idea of dropout was inspired from the benefit of ensemble learning in machine learning algorithms. Since the training and then combining multiple neural networks is not really possible in practice, the dropout algorithm attempts at creating virtual neural networks inside the main one by dropping some neurons in each iteration.


Dropout is a technique applied during the training state. For each update iteration, some of the neurons are removed randomly along with all their incoming and outgoing connections to get a thinned network. This can be done by randomly deciding if each neuron will be present before each iteration. Each node is present in the training phase with probability $p$  with the default value equal to 0.5.  Only those weights of survived  neurons are updated in the respective iteration. During the test time, a single neural network should be used to represent the combination of all thinned models. The authors in \cite{srivastava2014dropout} suggested using a scaled-down versions of the trained weights. During the training, approximately only $p$ of the whole neurons is used. Simply using weights from all neurons in the final combination scales up the expected output by $1/p$. Thus, a scaling factor $p$ needs to be multiplied to every weight when assembling all the thinned models. Dropout is very attractive due to its simplicity and strong regularization effect. Gradient descent variants such as Adam, momentum, other regularization methods and early stopping are also compatible with dropout.
  
\section{Transfer Learning and Data Augmentation}
\label{sec:transfer}
Another main problem in fingerprinting algorithms is the necessity of regular updates of fingerprinting database. Due to constant changes in indoor environments, the propagation environment changes with time and so do the fingerprints. It is necessary to regularly update the fingerprints or the model used for localization. 

Euclidean distance-based methods are instance-based learning methods which means that the training data is stored and every time a test point needs to be localized the whole database is used for localization. If a training data is out of date due to environment changes, the performance of fingerprinting algorithms is degraded and the solution is usually to abandon all the old data and collect new measurements for all points.  The need for regular updates of the database creates a burden for localization systems and is in general very time consuming. The problem remains for \ac{SVM} based solutions. 

In most of the practical situations, the general structure of the building remains intact. Intuitively if a learning algorithm is capable of implicitly learning the building structure then the model only needs to be fine tuned with slight changes of algorithm. Such a model can also be used as a pre-trained model for another building with similar structures. If buildings are similar in their indoor structure then the model trained for one of them can be fine-tuned to another one with small efforts. 

In deep learning research this is called transfer learning.  Deep learning models are shown to be capable of transferring the learning algorithm to different tasks. In image classification tasks, it has been shown that one can use a pre-trained model keeping the weights and the network size for a different dataset. Deep learning models like AlexNet, VGG and ResNet can be used as basis for new classification tasks \cite{razavian_cnn_2014,yosinski_how_2014}. The success of transfer learning is explained, as in  \cite{yosinski_how_2014}, by the feature extraction capabilities of deep neural networks. The first layer of deep neural networks in image classification tasks aims at extracting features that resemble either Gabor filters or color blobs. The feature extraction parts of these neural networks can still be used for other image datasets with different classification at hand only with additional fine tuning.

In this paper, the transfer learning is used for facilitating the fingerprinting in similar environments or for updating the database when the environment has been only slightly changed. The main assumption is that fingerprints can be used to learn an implicit representation of the building structure and this can be done by using deep neural networks. In the next section, we evaluate this idea by using a pre-trained model for a building and  update the model by a small amount of new data using standard gradient-based methods instead of training the network from the beginning. By doing so,  not only one can  take full use of the outdated data but also one can accelerate the whole training process.

\subsection{Data Augmentation}

In image recognition tasks, the output of classifiers should not be changed if the pictures are slightly transformed for example with small rotations. Accordingly a given training database can be enlarged by adding these transformations to the training set. In that way the learning algorithm is encouraged to learn those features essential to the classification task. 
The method is called data augmentation and  can be used to alleviate the effect of overfitting and compensate the lack of sufficient training samples.

In fingerprinting algorithms, due to multi-path effects on RSSI values, it is important to collect multiple measurements at the same point which in turn increases the collection time of training samples. However, if the RSSI values are obtained over time spans bigger than the coherence time of the channel, they can be seen as independent observations. In this case, the fingerprinting algorithm should not be sensitive to the order of fingerprints and therefore a permuted version of fingerprint vectors can also be used as corresponding to a location. 

In this paper, the training database is augmented using permuted version of RSSI values. Consider an $m\times k$ RSSI matrix where $m$ is the number of access points and $k$ is the number of measurements per access point. By permutation of each row independently, a new matrix with the same size is obtained. Permutation can be done multiple times for a dataset to further augment the data. As we will see in the next section, this technique leads to performance improvements for deep learning based fingerprinting localization.
%
%
\section{Numerical and Experimental Analysis}
\label{sec:Numerical}

In this section, the previous learning algorithms are implemented to solve indoor localization problem. The implementation details are discussed and the final algorithms are evaluated.

\subsection{Fingerprinting database}
\subsubsection{Simulation Data}
The propagation model used for simulating data in this work is present in \cite{borrelli2004channel}. The authors propose a multi-wall path loss model by analyzing the effect of number of walls on experimental data and the model can be written as:
\begin{equation}
L(d) = l_0+10\gamma \log(d)+l_c+kl_w \qquad(dB),
\end{equation}
where $l_0$ is a constant equal 40.22 dB for a center frequency of 2.45 GHz.  $\gamma$ is the path loss exponent, $d$ is the distance between transmitter and receiver and $k$ is the number of walls. $l_c$ is a constant for multi-wall loss model and $l_w$ is the wall attenuation factor.  $\gamma = 1.64$, $l_c=53.73$, $l_w=4.51$, $k=10$ are chosen to generate simulation data in this work. Apart from that, a random term $X$ drawn from exponential distribution $X\sim \mathrm{Exp}(\lambda)$ with $\lambda = 0.5$ has to be subtracted to model multipath propagation. Therefore the final model of RSSI values can be expressed as:
\begin{equation}
\text{RSSI}(d) = P_{tx}-(l_0+10\gamma \log(d)+l_c+kl_w )-X,
\end{equation}
where $P_{tx}$ is the transmit power of APs and are set to 20 dBm for all APs. The test environment is a $20 m\times 10 m$  room with four APs located at each corner. Five measurements per \ac{AP} is obtained for localization. The training grid consists of  a square lattice with 1m-length as size of each small square sides. In total there are 200 training points and 5 measurements are taken at each point. 1000 test points are drawn randomly in the room and their corresponding fingerprints are created similarly as training points. 

\subsubsection{Telecommunication Networks Group Data}
\begin{figure}[h]
	\centering
	\begin{subfigure}{.5\textwidth}
		\centering
		\includegraphics[width=.9\linewidth]{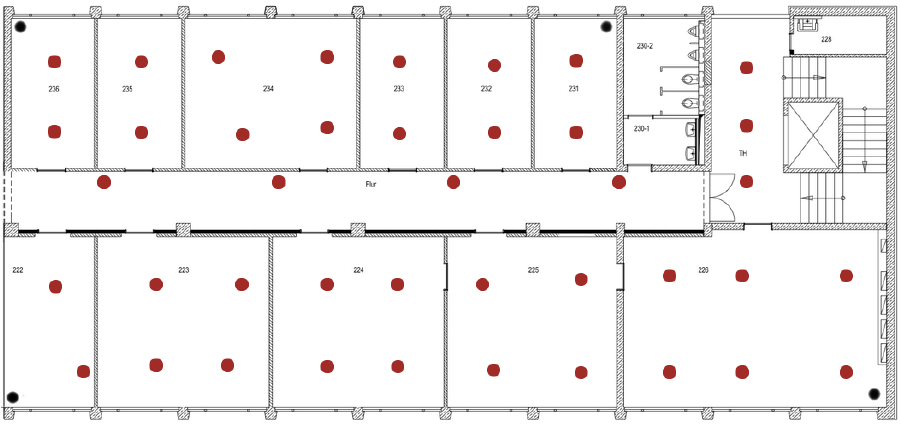}
		\caption{Training grid of TKN data}
	\end{subfigure}%
	\begin{subfigure}{.5\textwidth}
		\centering
		\includegraphics[width=.9\linewidth]{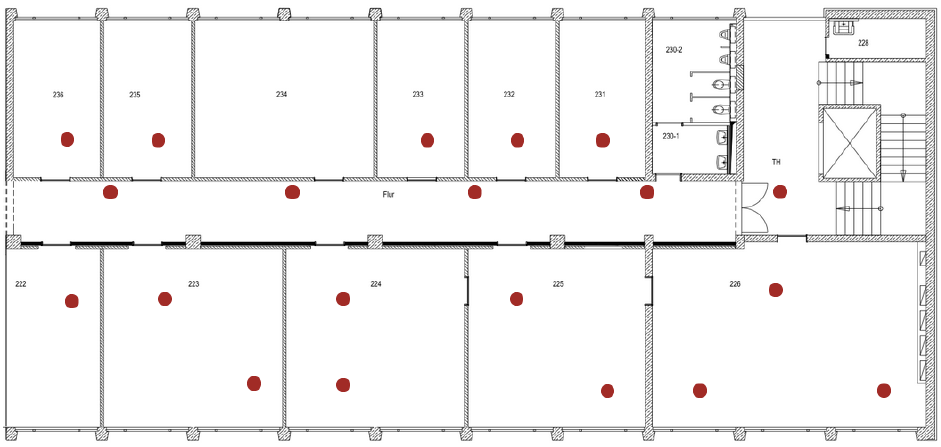}
		\caption{Testing grid of TKN data}
	\end{subfigure}
	\caption{Training and testing grids of TKN data \cite{lemic2014experimental}}
	\label{fig:TKN data}
\end{figure}

The Telecommunication Networks Group (TKN) data \cite{lemic2014demo} were collected in the TKN building in Berlin. In their work, fingerprints are constructed in different scenarios. The data in the scenario of "Small size office environment" is used for simulation in this work. The size of the whole area is approximately $30\times 15$ m.
\\
There are in total 116 APs deployed in the building for testing and the neighboring buildings. The dataset consists of 41 training points and 20 test points. The training and testing points are presented by red dots in Figure \ref{fig:TKN data}. 
\\
The TKN data can be obtained by sending request to the cloud services as described in \cite{lemicdata}. The raw measurements for one data point include many measurements from different APs. However, the number of measurements from different APs are not the same in TKN data due to the undetected signals. For instance, $AP_1$ may have 10 measurements while $AP_2$ just have 2. 

For our implementation, we use random sampling to address this problem. When one training sample is to be constructed, the RSSI value for a certain AP is sampled randomly from all available measurements of that AP at that point. At the end, 410 training samples and 20 test samples are constructed for the fingerprinting solutions.

\subsubsection{UJIIndoorLoc dataset}
In order to compare different proposed algorithms,  the UJIIndoorLoc dataset is also used \cite{torres2014ujiindoorloc}. It contains WiFi measurements used during EvAAL competition at IPIN
2015 \cite{moreira2015wi}. The dataset contains 19937 training samples and 1111 test samples. Each sample consists of 529 features where the first 520 features are \ac{RSSI} values from 520 access points ranged from -104 dBm to 0 dBm. The positive value 100 is used to indicate when a signal was not detected. The features 521 to 529 correspond to latitude, longitude, floor, building ID, space ID, relative position, user ID, phone ID and time stamp. The data in Table \ref{distribution} are obtained from 3 buildings with 3,3 and 4 floors respectively. 

\begin{table}[]
\centering
\caption{Distribution of data}
\label{distribution}
\begin{tabular}{|c|c|c|}
\hline
BuildingFloor ID & Training samples & Test samples \\ \hline\hline
B0F0               & 1059             & 78           \\ \hline
B0F1               & 1356             & 208          \\ \hline
B0F2               & 1443             & 165          \\ \hline
B0F3               & 1391             & 85           \\ \hline
B1F0               & 1368             & 30           \\ \hline
B1F1               & 1484             & 143          \\ \hline
B1F2               & 1396             & 87           \\ \hline
B1F3               & 948              & 47           \\ \hline
B2F0               & 1942             & 24           \\ \hline
B2F1               & 2162             & 111          \\ \hline
B2F2               & 1577             & 54           \\ \hline
B2F3               & 2709             & 40           \\ \hline
B2F4               & 1102             & 39           \\ \hline\hline
Total              & 19937            & 1111         \\ \hline
\end{tabular}
\end{table}
 
In this paper, we only select data from floors of building 0. The undetected signals are denoted by -110 dBm instead of 100 which means very weak signals (-104 dBm is the smallest measured RSSI value in data set). The features are then scaled independently to have zero mean and unit variance. The absolute positions are converted to relative positions by subtracting the smallest latitude and longitude in the data set. The room size (98.7m $\times$ 110.5m, 104.2m $\times$ 118.4m, 104.2m $\times$ 119.1m, 104.2m $\times$ 119.1m for four floors) can be obtained by looking at the difference between maximum and minimum of latitude and longitude. Moreover there are some access points that are undetected for all points in a certain floor. Those features are removed in order to speed up the training phase.

\subsection{Fingerprinting Algorithm Design}
In this work, we train a neural network as regression to learn the localization function mapping. The input layer corresponds to the \ac{RSSI} measurements with the number of neurons corresponding to the number of measurements. The output layer gives coordinates of a point in two dimensional space.  
The input dimension is at least as large as number of \acp{AP} in the environment. However in this work multiple measurements per \ac{AP} is included in the input. 
As it was discussed above, random sampling is used when the number of measurements from an \ac{AP} is not sufficient to give an input. It will be discussed below how this technique can be used for data augmentation. If no measurement is available from an \ac{AP}, the input values are set to the smallest possible \ac{RSSI} value in WiFi standard. The input layer size  is 20, that is five measurements per \ac{AP}, for the simulation data, 116 for TKN dataset and 520 for UJIIndoorLoc dataset.

\begin{table}[h]
\centering
\caption{Neural network configuration}
\label{tbl:nndesign}
\begin{tabular}{||c c c c c c||}
 \hline
\multicolumn{6}{||c||}{Raw RSS values as input}                \\ \hline\hline
\multicolumn{6}{||c||}{Fully Connected layer with 500 neurons} \\ \hline
\multicolumn{6}{||c||}{Dropout layer with 50\% rate}           \\ \hline
\multicolumn{6}{||c||}{Fully Connected layer with 500 neurons} \\ \hline
\multicolumn{6}{||c||}{Dropout layer with 50\% rate}           \\ \hline
\multicolumn{6}{||c||}{Fully Connected layer with 500 neurons} \\ \hline
\multicolumn{6}{||c||}{Dropout layer with 50\% rate}           \\ \hline\hline
\multicolumn{6}{||c||}{Location coordinates (2 dimensional)}   \\ \hline
\end{tabular}
\end{table}

Deep learning architecture for learning localization functions consists of three fully connected hidden layers  with  500 neurons in each hidden layer. 
All hidden layers are equipped with the \ac{ReLU} non-linearity. The output layer is a linear layer. For each layer we deploy a dropout layer with dropping rate 50 percent. The weights are initialized by using random procedure suggested above. The neural network is trained using Adam algorithm with learning rate 0.001, momentum parameter 0.9 and mini-batch size 100. Moreover $\ell_2$ penalty is also used with the penalty parameter $\lambda$ set to 0.03. The details can be found in Table \ref{tbl:nndesign}. The regression network is benchmarked with Euclidean distance based fingerprinting and \ac{SVM} based methods. 

\subsection{Autoencoder design for Feature Extraction}
The test for using autoencoder as feature extraction is done on the simulation data. First an autoencoder is trained properly and then the encoder is used to transform the original inputs into another feature space. The Euclidean distance-based method and SVM are then used for the transformed data.
\subsubsection{System Architecture}
A single layer autoencoder with size 5 is adopted in the experiment. \ac{BGD} is used as updating algorithm with learning rate 1, batch size 50. The input is a vector of size 20 meaning 5 measurements per 4 APs. After applying the encoder to the original data, each sample is transformed into a 5-dimensional vector. The goal is to find a better representation of the data or in other words find an efficient fingerprint construction. 

Euclidean distance-based method and SVM are then used for the 5-D vectors. $k$ is chosen equal to $3$ for the number of $k-$nearest neighbors for both cases. SVM with RBF kernel is adopted with $\gamma=1/4$ and the penalty parameter $C=1$.

\subsubsection{Simulation Data Performance}
The performance of autoencoder feature extraction is compared with the case where the fingerprints are constructed by simple averaging of \ac{RSSI} values of each \ac{AP}. Euclidean distance-based (ED-based) method and SVM are used for pattern matching in both cases. 
\\
Localization error can be seen as the performance metric of algorithms which is defined as the Euclidean distance between the estimated position and the ground truth position. 

The results are shown in Table \ref{table:autoencodersimulationdata}. Using autoencoder to extract features of data and then applying ED-based method or SVM does not give better performance than doing it directly on the original data. There can be two different reasons for this problem. First, it might be that the autoencoder is not capable of extracting properly all the data features that are suitable for localization tasks. But on the other hand, it might be the case that the autoencoder extracts too much information for the localization tasks. In fingerprinting contexts, the second reason seems more plausible. Note that the simple averaging of \ac{RSSI} values provides a better performance despite the fact that average values are insufficient in general to represent a dataset. This surprising observation indicates that in some applications, not all the features are relevant for the task at hand. As mentioned above, batch methods for training autoencoders tend to learn the average value of the data which is not suitable for most of pattern recognition tasks. This is not the case for localization and surprisingly these methods might provide better features for fingerprinting. In any case, indoor localization applications seem not to rely on very complex features of the dataset and therefore no sophisticated feature extraction method is required.

\begin{table}[h]
	\centering
	\caption{Summary results for simulation data when applied feature extraction}
	\label{table:autoencodersimulationdata}
	\begin{tabular}{lllll}
		\hline
		\multicolumn{5}{c}{\textbf{Simulation Data}}                                                                                                 \\ \hline
		\multicolumn{1}{c}{}                       & \multicolumn{1}{c}{ED} & \multicolumn{1}{c}{SVM} & \multicolumn{1}{c}{Autoencoder+ED} 
		 & \multicolumn{1}{c}{Autoencoder+SVM}\\ \hline
		\multicolumn{1}{c}{Mean error {[}m{]}}     & \multicolumn{1}{c}{2.44}                   & \multicolumn{1}{c}{2.37}    & 
		\multicolumn{1}{c}{2.66}  & \multicolumn{1}{c}{2.47}                 
		\\ \hline
		\multicolumn{1}{c}{Error variance } & \multicolumn{1}{c}{2.27}                   & \multicolumn{1}{c}{2.19}    & \multicolumn{1}{c}{2.39}               & \multicolumn{1}{c}{2.49}           
		\\ \hline
		\multicolumn{1}{c}{Min. error {[}m{]}}     & \multicolumn{1}{c}{0.04}                   & \multicolumn{1}{c}{0.03}    & \multicolumn{1}{c}{0.11}               & \multicolumn{1}{c}{0.12}    
		\\ \hline
		\multicolumn{1}{c}{Max. error {[}m{]}}     & \multicolumn{1}{c}{10.63}                   & \multicolumn{1}{c}{8.33}    & \multicolumn{1}{c}{7.90}              & \multicolumn{1}{c}{8.76}     \\ \hline
	\end{tabular}
\end{table}
\subsection{Performance of Regression Networks in Indoor Localization}
\subsubsection{Simulation Data Performance}
\begin{figure}[h]
	\centerline{\includegraphics[width=0.65\textwidth]{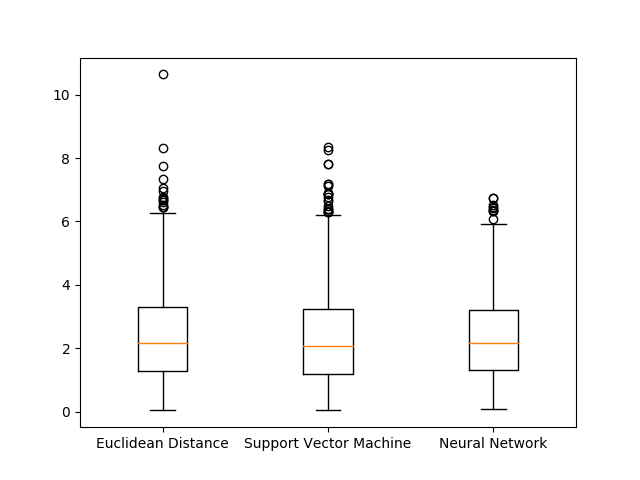}}
	\caption{\scriptsize Box plots of localization errors of three algorithms for simulation data}
	\label{fig:SimulationComparison}
\end{figure}
\begin{table}[h]
	\centering
	\caption{Summary results of three algorithms for simulation data}
	\label{table:simulationdata}
	\begin{tabular}{llll}
		\hline
		\multicolumn{4}{c}{\textbf{Simulation Data}}                                                                                                 \\ \hline
		\multicolumn{1}{c}{}                       & \multicolumn{1}{c}{Euclidean Distance} & \multicolumn{1}{c}{Support Vector Machine} & \multicolumn{1}{c}{Neural Network} \\ \hline
		\multicolumn{1}{c}{Mean error {[}m{]}}     & \multicolumn{1}{c}{2.44}                   & \multicolumn{1}{c}{2.37}    & \multicolumn{1}{c}{2.35}               \\ \hline
		\multicolumn{1}{c}{Error variance } & \multicolumn{1}{c}{2.27}                   & \multicolumn{1}{c}{2.19}    & \multicolumn{1}{c}{1.75}               \\ \hline
		\multicolumn{1}{c}{Min. error {[}m{]}}     & \multicolumn{1}{c}{0.04}                   & \multicolumn{1}{c}{0.03}    & \multicolumn{1}{c}{0.06}               \\ \hline
		\multicolumn{1}{c}{Max. error {[}m{]}}     & \multicolumn{1}{c}{10.63}                   & \multicolumn{1}{c}{8.33}    & \multicolumn{1}{c}{6.73}               \\ \hline
	\end{tabular}
\end{table}
As shown in Table \ref{table:simulationdata}, the neural network approach has the least mean error with 2.35 meters. Moreover, it can be seen in Figure \ref{fig:SimulationComparison} that there are less outliers for the neural network approach than the Euclidean distance-based or SVM approach. This is also proven by comparing the variance of error for the neural network compared to other algorithms. For the simulation data, the neural network approach not only gives a better localization result but also a more stable performance. 

\subsubsection{UJIndoorLoc Data Performance}

\begin{table}[h]
	\centering
	\caption{Summary results of three algorithms for UJIndoorLoc Data}
	\label{UJIndoorLoc Data result}
	\begin{tabular}{llll}
		\hline
 		\multicolumn{1}{c}{}                       & \multicolumn{1}{c}{Euclidean Distance} & \multicolumn{1}{c}{Support Vector Machine} & \multicolumn{1}{c}{Neural Network}                                                                                               \\ \hline
 				\multicolumn{4}{c}{\textbf{Floor 0 of Building 0}} 
 		\\ \hline
		\multicolumn{1}{c}{Mean error {[}m{]}}     & \multicolumn{1}{c}{10.06}                   & \multicolumn{1}{c}{8.48}    & \multicolumn{1}{c}{7.62}               \\ \hline
		\multicolumn{1}{c}{Error variance } & \multicolumn{1}{c}{68.30}                   & \multicolumn{1}{c}{63.35}    & \multicolumn{1}{c}{43.00}               \\ \hline
		\multicolumn{1}{c}{Min. error {[}m{]}}     & \multicolumn{1}{c}{0.45}                   & \multicolumn{1}{c}{0.25}    & \multicolumn{1}{c}{0.63}               \\ \hline
		\multicolumn{1}{c}{Max. error {[}m{]}}     & \multicolumn{1}{c}{40.15}                   & \multicolumn{1}{c}{53.62}    & \multicolumn{1}{c}{34.02}               \\ \hline
		
		\multicolumn{4}{c}{\textbf{Floor 1 of Building 0}}                                                                                                 \\ \hline
		Mean error {[}m{]}                         &         \multicolumn{1}{c}{9.93}                           &     \multicolumn{1}{c}{8.81}              & \multicolumn{1}{c}{8.08}                             \\ \hline
		Error variance                    &     \multicolumn{1}{c}{195.33}                                   &       \multicolumn{1}{c}{119.73}                   &                  \multicolumn{1}{c}{93.94}                   \\ \hline
		Min. error {[}m{]}                         &\multicolumn{1}{c}{0.25}                                        &\multicolumn{1}{c}{0.09}                          &\multicolumn{1}{c}{0.40}                                     \\ \hline
		Max. error {[}m{]}                         &\multicolumn{1}{c}{118.64}                                        &\multicolumn{1}{c}{81.56}                        & \multicolumn{1}{c}{90.00}                                   \\ \hline
		
		\multicolumn{4}{c}{\textbf{Floor 2 of Building 0}}                                                                                                 \\ \hline
	
		Mean error {[}m{]}                         &         \multicolumn{1}{c}{9.50}                           &     \multicolumn{1}{c}{9.42}              & \multicolumn{1}{c}{7.42}                             \\ \hline
		Error variance                      &     \multicolumn{1}{c}{180.24}                                   &       \multicolumn{1}{c}{175.47}                   &                  \multicolumn{1}{c}{28.43}                   \\ \hline
		Min. error {[}m{]}                         &\multicolumn{1}{c}{0.07}                                        &\multicolumn{1}{c}{0.38}                          &\multicolumn{1}{c}{0.37}                                     \\ \hline
		Max. error {[}m{]}                         &\multicolumn{1}{c}{105.58}                                        &\multicolumn{1}{c}{86.69}                        & \multicolumn{1}{c}{25.21}                                   \\ \hline
		
		\multicolumn{4}{c}{\textbf{Floor 3 of Building 0}}                                                                                                 \\ \hline
		Mean error {[}m{]}                         &         \multicolumn{1}{c}{9.40}                           &     \multicolumn{1}{c}{7.72}              & \multicolumn{1}{c}{7.27}                             \\ \hline
		Error variance                     &     \multicolumn{1}{c}{89.87}                                   &       \multicolumn{1}{c}{52.98}                   &                  \multicolumn{1}{c}{29.36}                   \\ \hline
		Min. error {[}m{]}                         &\multicolumn{1}{c}{0.82}                                        &\multicolumn{1}{c}{0.22}                          &\multicolumn{1}{c}{0.76}                                     \\ \hline
		Max. error {[}m{]}                         &\multicolumn{1}{c}{63.35}                                        &\multicolumn{1}{c}{39.16}                        & \multicolumn{1}{c}{26.41}                                   \\ \hline
	\end{tabular}
\end{table}
For simplicity, the evaluation is only limited to the building 0 of UJIndoorLoc data. Since there are no available validation data, the training data are randomly divided into training data and validation data with 70\% and 30\%. In Table \ref{UJIndoorLoc Data result}, performances of three algorithms for UJIndoorLoc data are compared. The results are similar to the simulation data. The neural network approach has a smaller mean localization error. Although it does not always have the lowest minimum and maximum errors, the variance is the smallest which indicates more stable localization ability. Notice that in this case, there are 520 APs compared to 4 APs in the simulation data. This also proves the scalability  of neural networks when there are lots of features. 


\subsubsection{TKN Data Performance}
\begin{table}[h]
	\centering
	\caption{Summary results of three algorithms for TKN data}
	\label{table:TKNdatacomparison}
	\begin{tabular}{llll}
		\hline
		\multicolumn{4}{c}{\textbf{TKN Data}}                                                                                                 \\ \hline
		\multicolumn{1}{c}{}                       & \multicolumn{1}{c}{Euclidean Distance} & \multicolumn{1}{c}{Support Vector Machine} & \multicolumn{1}{c}{Neural Network} \\ \hline
		\multicolumn{1}{c}{Mean error {[}m{]}}     & \multicolumn{1}{c}{4.90}                   & \multicolumn{1}{c}{5.05}    & \multicolumn{1}{c}{3.54}               \\ \hline
		\multicolumn{1}{c}{Error variance } & \multicolumn{1}{c}{9.02}                   & \multicolumn{1}{c}{11.97}    & \multicolumn{1}{c}{5.97}               \\ \hline
		\multicolumn{1}{c}{Min. error {[}m{]}}     & \multicolumn{1}{c}{0.00}                   & \multicolumn{1}{c}{1.22}    & \multicolumn{1}{c}{0.59}               \\ \hline
		\multicolumn{1}{c}{Max. error {[}m{]}}     & \multicolumn{1}{c}{9.97}                   & \multicolumn{1}{c}{14.62}    & \multicolumn{1}{c}{9.63}               \\ \hline
	\end{tabular}
\end{table}
In TKN data, there are a lot of unimportant features (APs in the other buildings) which is a challenge for localization algorithms. The results in Table \ref{table:TKNdatacomparison} show the ability of deep learning in dealing with irrelevant features. Neural network gives much better localization accuracy than Euclidean distance-based and SVM methods. The neural network approach also gives smaller variance, the same as shown in the previous two datasets. 
\subsection{Performance of Data Augmentation}
The test of data augmentation is done on both simulation data and UJIndoorLoc data. The data is augmented 1, 5 and 10 times to compare the influence of different levels of data augmentation. One time augmentation means that the dataset is two times as large as the original dataset. Notice that there are 4 APs for the simulation data while 520 APs for UJIndoorLoc data. The test for UJIndoorLoc data is only done for the building 0. Similarly to the previous evaluation, average error, error variance, minimum and maximum error are the performance metrics.
\subsubsection{Simulation Data Performance}
The simulation data contains 1000 training points and 1000 test points. There are 2000, 5000, 11000 training points after 1, 5, 10 times augmentation respectively. The localization performance is still evaluated by 1000 test points.
\begin{table}[h]
	\centering
	\caption{Summary results for simulation data when applied data augmentation}
	\label{table:dataaugmentation}
	\begin{tabular}{lllll}
		\hline
		\multicolumn{5}{c}{\textbf{Simulation Data}}                                                                                                 \\ \hline
		\multicolumn{1}{c}{}                       & \multicolumn{1}{c}{Original} & \multicolumn{1}{c}{1 time permutation} & \multicolumn{1}{c}{5 times permutation} & \multicolumn{1}{c}{10 times permutation}\\ \hline
		\multicolumn{1}{c}{Mean error {[}m{]}}     & \multicolumn{1}{c}{2.35}                   & \multicolumn{1}{c}{2.33}    & \multicolumn{1}{c}{2.23}               
		&\multicolumn{1}{c}{2.21}
		\\ \hline
		\multicolumn{1}{c}{Error variance } & \multicolumn{1}{c}{1.75}                   & \multicolumn{1}{c}{1.73}    & \multicolumn{1}{c}{1.59}               
		&\multicolumn{1}{c}{1.56}
		\\ \hline
		\multicolumn{1}{c}{Min. error {[}m{]}}     & \multicolumn{1}{c}{0.06}                   & \multicolumn{1}{c}{0.04}    & \multicolumn{1}{c}{0.05}               
		&\multicolumn{1}{c}{0.02}
		\\ \hline
		\multicolumn{1}{c}{Max. error {[}m{]}}     & \multicolumn{1}{c}{6.73}                   & \multicolumn{1}{c}{8.01}    & \multicolumn{1}{c}{8.07}               
		&\multicolumn{1}{c}{6.93}
		\\ \hline
	\end{tabular}
\end{table}
\\
By comparing the mean localization error in Table \ref{table:dataaugmentation}, the improvement of data augmentation for fingerprinting indoor localization approach can be observed. Notice that with the increase of augmentation level, the localization error is also decreasing. By 5 times permuting the fingerprints, the localization error is decreased by about 10\%. However, with 10 times fingerprints permutation, the localization error is almost the same as the error of doing 5 times permutation. It seems that at a certain level of data augmentation, the performance becomes saturated. 

\subsubsection{UJIndoorLoc Data Performance}
The test of data augmentation for experimental data is done only for floor 0 of building 0 in UJIndoorLoc data consisting of 1059 training samples and 78 test samples. After data augmentation, 30\% of the whole data are randomly picked as validation data.
\\
The results in Table \ref{table:dataaugmentationUJI} verify the effectiveness of data augmentation for experimental data. 10 times permutation of fingerprints decrease the average localization error by about 1 meter as well as the error variance. Similarly to simulation data, the improvement of performance is getting lower when the augmentation reaches a certain level.
\begin{table}[h]
	\centering
	\caption{Summary results for UJIndoorLoc data  when applied data augmentation}
	\label{table:dataaugmentationUJI}
	\begin{tabular}{lllll}
		\hline
		\multicolumn{5}{c}{\textbf{Floor 0 of Building 0}}                                                                                                 \\ \hline
		\multicolumn{1}{c}{}                       & \multicolumn{1}{c}{Original} & \multicolumn{1}{c}{1 time permutation} & \multicolumn{1}{c}{5 times permutation} & \multicolumn{1}{c}{10 times permutation}\\ \hline
		\multicolumn{1}{c}{Mean error {[}m{]}}     & \multicolumn{1}{c}{7.62}                   & \multicolumn{1}{c}{6.99}    & \multicolumn{1}{c}{6.87}               
		&\multicolumn{1}{c}{6.69}
		\\ \hline
		\multicolumn{1}{c}{Error variance } & \multicolumn{1}{c}{43.00}                   & \multicolumn{1}{c}{41.48}    & \multicolumn{1}{c}{40.85}               
		&\multicolumn{1}{c}{38.72}
		\\ \hline
		\multicolumn{1}{c}{Min. error {[}m{]}}     & \multicolumn{1}{c}{0.63}                   & \multicolumn{1}{c}{0.20}    & \multicolumn{1}{c}{0.37}               
		&\multicolumn{1}{c}{0.45}
		\\ \hline
		\multicolumn{1}{c}{Max. error {[}m{]}}     & \multicolumn{1}{c}{34.02}                   & \multicolumn{1}{c}{34.16}    & \multicolumn{1}{c}{31.02}               
		&\multicolumn{1}{c}{30.17}
		\\ \hline
	\end{tabular}
\end{table}
\\
The performance improvement by data augmentation is almost free. When there is enough memory and computation power, choosing the number of permutation large enough is a good choice. Although the training time may increase due to more training samples but this computation is offline and the localization of test points are still done by a single forward propagation and thus can be used in real applications.
\subsection{Performance of Transfer Learning}
\begin{figure}[h]
	\centerline{\includegraphics[width=0.7\textwidth]{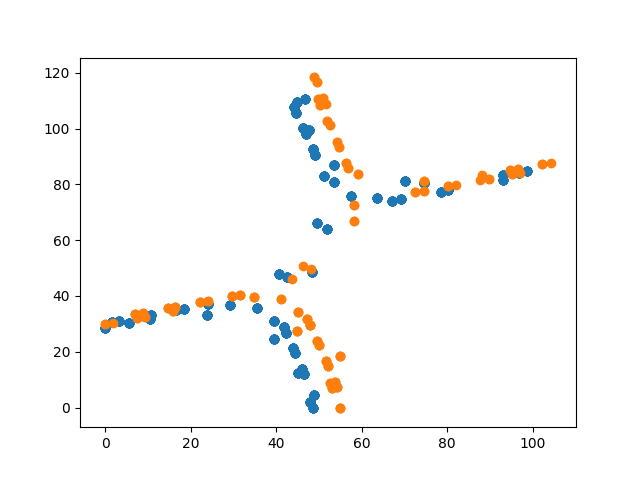}}
	\caption{\scriptsize Training grids of floor 0 and 1 of building 0 in UJIndoorLoc data. It can be seen that the training grids are following the same track indicating a similar inner structure of two floors.}
	\label{fig:floorcomparison}
\end{figure}

In order to test the performance of transfer learning, it is necessary to have data from two different propagation models. Two different floors of one building in UJIndoorLoc data can be seen as two different propagation models. They share the same structure as can be seen from Figure \ref{fig:floorcomparison} with similar anchor placement and thus can be used to test the usage of transfer learning.
First, the neural network is trained on the whole dataset of floor 0 to get the pre-trained model and then the model is fine-tuned by only 30\% of the training data from floor 1. At the end the model is tested on the test data of floor 1. 
\begin{figure}[h]
	\centerline{\includegraphics[width=0.7\textwidth]{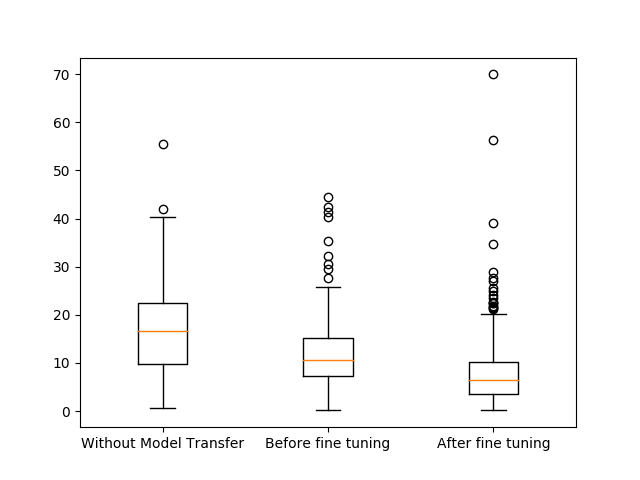}}
	\caption{\scriptsize Box plots of localization errors for transfer learning}
	\label{fig:ModelTransfer}
\end{figure}
\\
The test results shown in Table \ref{modeltransfer} include three situations. First, the neural network is trained directly by 30\% of the training data from floor 1 and tested with test samples from floor 1. This shows whether smaller training data is sufficient or not. Secondly, the neural network is trained by the whole data from floor 0 and tested with the test samples from floor 1. This experiment examines the performance of naive transfer learning. Lastly, the model obtained from the floor 0 is fine-tuned by 30\% of the training data from floor 1 and tested with the test samples from floor 1. The average error after fine-tuning shows a comparable mean error compared with using the whole training data. The result shows the potential of using transfer learning in fingerprinting localization. By comparing the first and third situation, it can be seen that even when there are not enough data, using a pre-trained model can help to get a reasonable performance. The second and third situation can be seen as a model update. The old model (floor 0) can be updated by a small amount of new data (floor 1) and it will work well in the new environment.
\\
\begin{table}[h]
	\centering
	\caption{Summary results for transfer learning}
	\label{modeltransfer}
	\vspace*{-0.5cm}
	\hspace*{-0.3cm}
	\begin{tabular}{llll}    
		\\                                          		\hline
		\multicolumn{4}{c}{\textbf{Floor 0, 1 of Building 0}}                                                                                                 \\ \hline
		\multicolumn{1}{c}{}                       & \multicolumn{1}{c}{Without transfer learning} & \multicolumn{1}{c}{Before fine-tuning} & \multicolumn{1}{c}{After fine-tuning} \\ \hline
		\multicolumn{1}{c}{Mean error {[}m{]}}     & \multicolumn{1}{c}{16.76}                   & \multicolumn{1}{c}{11.90}    & \multicolumn{1}{c}{8.70}               \\ \hline
		\multicolumn{1}{c}{Error variance } & \multicolumn{1}{c}{77.75}                   & \multicolumn{1}{c}{54.75}    & \multicolumn{1}{c}{73.24}               \\ \hline
		\multicolumn{1}{c}{Min. error {[}m{]}}     & \multicolumn{1}{c}{0.67}                   & \multicolumn{1}{c}{0.21}    & \multicolumn{1}{c}{0.15}               \\ \hline
		\multicolumn{1}{c}{Max. error {[}m{]}}     & \multicolumn{1}{c}{55.45}                   & \multicolumn{1}{c}{44.53}    & \multicolumn{1}{c}{69.93}               \\ \hline
	\end{tabular}
\end{table}
 

\section{Conclusion} 
The main motivation behind this work is to show how artificial intelligence can be utilized for indoor localization applications to accomplish tasks that cannot be done efficiently by the existing methods. As expected a complex neural network architecture is capable of approximating well the localization function mapping the measured \ac{RSSI} values to the locations. Moreover we proposed using data augmentation and transfer learning to alleviate the problem of data collection which is essential in fingerprinting approaches. The structure of a particular environment is reflected in the trained neural network. Therefore a neural network can provide reasonable performance when it is used in another building with same anchor placement. This can significantly simplify the way in which fingerprinting algorithms are trained.

\section{Acknowledgment}
The authors are grateful to Filip Lemic for helping with TKN dataset.

\bibliographystyle{IEEEtran}
\bibliography{mybibliography}
 
\end{document}

%% file: acronym_def.tex
\acrodef{AP}{Access Points}
\acrodef{BGD}{Batch Gradient Descent}
\acrodef{CSI}{Channel State Information}
\acrodef{DNN}{Deep Neural Network}
\acrodef{ED}{Euclidean Distance}
\acrodef{FPS}{Fingerprinting Localization Solution}
\acrodef{KL}{Kullback-Leibler}
\acrodef{kNN}{k-nearest neighbor}
\acrodef{GCN}{Global contrast normalization} 
\acrodef{MBGD}{Mini-batch Gradient Descent}
\acrodef{PCA}{Principle Component Analysis}
\acrodef{RBF}{Radial Basis Function}
\acrodef{ReLU}{Rectified Linear Unit}
\acrodef{RF}{Radio Frequency}
\acrodef{RSSI}{Received Signal Strength Indicator}
\acrodef{SGD}{Stochastic Gradient Descent}
\acrodef{SNR}{Signal-to-Noise-Ratio}
\acrodef{SVM}{Support Vector Machine}

%% file: localization.bbl
\begin{thebibliography}{10}
\providecommand{\url}[1]{#1}
\csname url@samestyle\endcsname
\providecommand{\newblock}{\relax}
\providecommand{\bibinfo}[2]{#2}
\providecommand{\BIBentrySTDinterwordspacing}{\spaceskip=0pt\relax}
\providecommand{\BIBentryALTinterwordstretchfactor}{4}
\providecommand{\BIBentryALTinterwordspacing}{\spaceskip=\fontdimen2\font plus
\BIBentryALTinterwordstretchfactor\fontdimen3\font minus
  \fontdimen4\font\relax}
\providecommand{\BIBforeignlanguage}[2]{{%
\expandafter\ifx\csname l@#1\endcsname\relax
\typeout{** WARNING: IEEEtran.bst: No hyphenation pattern has been}%
\typeout{** loaded for the language `#1'. Using the pattern for}%
\typeout{** the default language instead.}%
\else
\language=\csname l@#1\endcsname
\fi
#2}}
\providecommand{\BIBdecl}{\relax}
\BIBdecl

\bibitem{medina_ultrasound_2013}
C.~Medina, J.~Segura, and Ã.~De~la Torre, ``\BIBforeignlanguage{en}{Ultrasound
  {Indoor} {Positioning} {System} {Based} on a {Low}-{Power} {Wireless}
  {Sensor} {Network} {Providing} {Sub}-{Centimeter} {Accuracy}},''
  \emph{\BIBforeignlanguage{en}{Sensors}}, vol.~13, no.~3, pp. 3501--3526, Mar.
  2013.

\bibitem{brassart_localization_2000}
E.~Brassart, C.~Pegard, and M.~Mouaddib, ``Localization using infrared
  beacons,'' \emph{Robotica}, vol.~18, no.~02, pp. 153--161, Mar. 2000.

\bibitem{erol-kantarci_survey_2011}
Erol-Kantarci \emph{et~al.}, ``A {Survey} of {Architectures} and {Localization}
  {Techniques} for {Underwater} {Acoustic} {Sensor} {Networks},'' \emph{IEEE
  Communications Surveys Tutorials}, vol.~13, no.~3, pp. 487--502, 2011.

\bibitem{amundson_survey_2009}
I.~Amundson and X.~D. Koutsoukos, ``\BIBforeignlanguage{en}{A {Survey} on
  {Localization} for {Mobile} {Wireless} {Sensor} {Networks}},'' in
  \emph{\BIBforeignlanguage{en}{Mobile {Entity} {Localization} and {Tracking}
  in {GPS}-less {Environnments}}}, ser. Lecture {Notes} in {Computer}
  {Science}, R.~Fuller and X.~D. Koutsoukos, Eds.\hskip 1em plus 0.5em minus
  0.4em\relax Springer Berlin Heidelberg, 2009, no. 5801, pp. 235--254.

\bibitem{Seco2009}
F.~Seco \emph{et~al.}, ``A survey of mathematical methods for indoor
  localization,'' in \emph{Intelligent Signal Processing}, 2009, pp. 9 --14.

\bibitem{Milioris2014}
D.~Milioris \emph{et~al.}, ``Low-dimensional signal-strength fingerprint-based
  positioning in wireless lans,'' \emph{Ad Hoc Networks}, pp. 100 -- 114, 2014.

\bibitem{honkavirta2009comparative}
V.~Honkavirta \emph{et~al.}, ``{A Comparative Survey of WLAN Location
  Fingerprinting Methods},'' in \emph{WPNC 2009}.\hskip 1em plus 0.5em minus
  0.4em\relax IEEE, 2009, pp. 243--251.

\bibitem{milioris2011low}
D.~Milioris \emph{et~al.}, ``{Low-Dimensional Signal-Strength Fingerprint-based
  Positioning in Wireless LANs},'' \emph{Ad Hoc Networks}, 2011.

\bibitem{Laoudias2009}
C.~Laoudias \emph{et~al.}, ``Localization using radial basis function networks
  and signal strength fingerprints in wlan,'' in \emph{Global
  Telecommunications Conference, 2009. GLOBECOM 2009. IEEE}, 2009, pp. 1--6.

\bibitem{Bai2013}
S.~Bai and T.~Wu, ``Analysis of k-means algorithm on fingerprint based indoor
  localization system,'' in \emph{Microwave, Antenna, Propagation and EMC
  Technologies for Wireless Communications}, 2013.

\bibitem{Steiner2011}
C.~Steiner and A.~Wittneben, ``Efficient training phase for ultrawideband-based
  location fingerprinting systems,'' \emph{Signal Processing, IEEE Transactions
  on}, vol.~59, no.~12, pp. 6021--6032, 2011.

\bibitem{Machaj2010}
J.~Machaj \emph{et~al.}, ``Impact of the number of access points in indoor
  fingerprinting localization,'' in \emph{Radioelektronika}, 2010, pp. 1--4.

\bibitem{Kaemarungsi2004}
K.~Kaemarungsi \emph{et~al.}, ``Modeling of indoor positioning systems based on
  location fingerprinting,'' in \emph{INFOCOM}, 2004, pp. 1012--1022.

\bibitem{Kaemarungsi2005}
K.~Kaemarungsi, ``Efficient design of indoor positioning systems based on
  location fingerprinting,'' in \emph{Wireless Networks, Communications and
  Mobile Computing}, vol.~1, 2005, pp. 181--186.

\bibitem{wen_fundamental_2015}
Y.~Wen, X.~Tian, X.~Wang, and S.~Lu, ``Fundamental limits of {RSS}
  fingerprinting based indoor localization,'' in \emph{2015 {IEEE} {Conference}
  on {Computer} {Communications} ({INFOCOM})}, Apr. 2015, pp. 2479--2487.

\bibitem{behboodi17hypothesis}
A.~Behboodi, F.~Lemic, and A.~Wolisz, ``{Hypothesis Testing Based Model for
  Fingerprinting Localization Algorithms},'' in \emph{2017 IEEE 85th Vehicular
  Technology Conference (VTC-Spring'17)}, 2017.

\bibitem{behboodi_mathematical_2016}
\BIBentryALTinterwordspacing
A.~Behboodi \emph{et~al.}, ``A {Mathematical} {Model} for
  {Fingerprinting}-based {Localization} {Algorithms},'' \emph{arXiv preprint},
  2016, arXiv: 1610.07636. [Online]. Available: \url{arxiv:1610.07636}
\BIBentrySTDinterwordspacing

\bibitem{BehboodiIPIN2017}
A.~Behboodi, F.~Lemic, A.~Wolisz, and R.~Mathar, ``Interference effect on the
  performance of fingerprinting localization,'' in \emph{International
  Conference on Indoor Positioning and Indoor Navigation (IPIN 2017)},
  September 2017.

\bibitem{Ding2013}
G.~Ding \emph{et~al.}, ``Overview of received signal strength based
  fingerprinting localization in indoor wireless lan environments,'' in
  \emph{Microwave, Antenna, Propagation and EMC Technologies for Wireless
  Communications}, 2013.

\bibitem{wang2017csi}
X.~Wang, L.~Gao, S.~Mao, and S.~Pandey, ``Csi-based fingerprinting for indoor
  localization: A deep learning approach,'' \emph{IEEE Transactions on
  Vehicular Technology}, vol.~66, no.~1, pp. 763--776, 2017.

\bibitem{nowicki2016low}
M.~Nowicki and J.~Wietrzykowski, ``Low-effort place recognition with wifi
  fingerprints using deep learning,'' \emph{arXiv preprint arXiv:1611.02049},
  2016.

\bibitem{Linchenetal2017}
L.~Xiao, A.~Behboodi, and R.~Mathar, ``A deep learning approach to
  fingerprinting indoor localization,'' in \emph{International
  Telecommunication Networks and Applications conference (ITNAC)}, November
  2017.

\bibitem{lemic2014experimental}
F.~Lemic, A.~Behboodi, V.~Handziski, and A.~Wolisz, ``Experimental
  decomposition of the performance of fingerprinting-based localization
  algorithms,'' in \emph{Indoor Positioning and Indoor Navigation (IPIN), 2014
  International Conference on}.\hskip 1em plus 0.5em minus 0.4em\relax IEEE,
  2014, pp. 355--364.

\bibitem{kandel_wrangler_2011}
S.~Kandel, A.~Paepcke, J.~Hellerstein, and J.~Heer, ``Wrangler: {Interactive}
  visual specification of data transformation scripts,'' in \emph{Proceedings
  of the {SIGCHI} {Conference} on {Human} {Factors} in {Computing}
  {Systems}}.\hskip 1em plus 0.5em minus 0.4em\relax ACM, 2011, pp. 3363--3372.

\bibitem{wickham_tidy_2014}
H.~Wickham and {others}, ``Tidy data,'' \emph{Journal of Statistical Software},
  vol.~59, no.~10, pp. 1--23, 2014.

\bibitem{codd_relational_1990}
E.~F. Codd, \emph{The {Relational} {Model} for {Database} {Management}:
  {Version} 2}.\hskip 1em plus 0.5em minus 0.4em\relax Boston, MA, USA:
  Addison-Wesley Longman Publishing Co., Inc., 1990.

\bibitem{cortes1995support}
C.~Cortes and V.~Vapnik, ``Support-vector networks,'' \emph{Machine learning},
  vol.~20, no.~3, pp. 273--297, 1995.

\bibitem{steinwart_support_2008}
I.~Steinwart and A.~Christmann, \emph{Support vector machines}, 1st~ed., ser.
  Information science and statistics.\hskip 1em plus 0.5em minus 0.4em\relax
  New York: Springer, 2008.

\bibitem{hsu2002comparison}
C.-W. Hsu and C.-J. Lin, ``A comparison of methods for multiclass support
  vector machines,'' \emph{IEEE transactions on Neural Networks}, vol.~13,
  no.~2, pp. 415--425, 2002.

\bibitem{minsky_perceptrons_1972}
M.~L. Minsky and S.~A. Papert, \emph{\BIBforeignlanguage{eng}{Perceptrons: an
  introduction to computational geometry}}, 2nd~ed.\hskip 1em plus 0.5em minus
  0.4em\relax Cambridge/Mass.: The MIT Press, 1972.

\bibitem{cybenko_approximation_1989}
G.~Cybenko, ``\BIBforeignlanguage{en}{Approximation by superpositions of a
  sigmoidal function},'' \emph{\BIBforeignlanguage{en}{Mathematics of Control,
  Signals and Systems}}, vol.~2, no.~4, pp. 303--314, Dec. 1989.

\bibitem{hornik_approximation_1991}
K.~Hornik, ``Approximation capabilities of multilayer feedforward networks,''
  \emph{Neural Networks}, vol.~4, no.~2, pp. 251--257, 1991.

\bibitem{sonoda_neural_2017}
S.~Sonoda and N.~Murata, ``Neural network with unbounded activation functions
  is universal approximator,'' \emph{Applied and Computational Harmonic
  Analysis}, vol.~43, no.~2, pp. 233--268, Sep. 2017.

\bibitem{raghu_expressive_2017}
M.~Raghu, B.~Poole, J.~Kleinberg, S.~Ganguli, and J.~Sohl-Dickstein,
  ``\BIBforeignlanguage{en}{On the {Expressive} {Power} of {Deep} {Neural}
  {Networks}},'' in \emph{\BIBforeignlanguage{en}{{PMLR}}}, Jul. 2017, pp.
  2847--2854.

\bibitem{bengio2012practical}
Y.~Bengio, ``Practical recommendations for gradient-based training of deep
  architectures,'' in \emph{Neural networks: Tricks of the trade}.\hskip 1em
  plus 0.5em minus 0.4em\relax Springer, 2012, pp. 437--478.

\bibitem{linsker_self_organization_1988}
R.~Linsker, ``Self-organization in a perceptual network,'' \emph{IEEE
  Computer}, vol.~21, no.~3, pp. 105--117, Mar. 1988.

\bibitem{bengio2007greedy}
Y.~Bengio, P.~Lamblin, D.~Popovici, H.~Larochelle \emph{et~al.}, ``Greedy
  layer-wise training of deep networks,'' \emph{Advances in neural information
  processing systems}, vol.~19, p. 153, 2007.

\bibitem{vincent2008extracting}
P.~Vincent, H.~Larochelle, Y.~Bengio, and P.-A. Manzagol, ``Extracting and
  composing robust features with denoising autoencoders,'' in \emph{Proceedings
  of the 25th international conference on Machine learning}.\hskip 1em plus
  0.5em minus 0.4em\relax ACM, 2008, pp. 1096--1103.

\bibitem{vincent2010stacked}
P.~Vincent, H.~Larochelle, I.~Lajoie, Y.~Bengio, and P.-A. Manzagol, ``Stacked
  denoising autoencoders: Learning useful representations in a deep network
  with a local denoising criterion,'' \emph{Journal of Machine Learning
  Research}, vol.~11, no. Dec, pp. 3371--3408, 2010.

\bibitem{leo_breiman_bagging_1996}
{Leo Breiman}, ``Bagging predictors,'' \emph{Machine Learning}, vol.~26, pp.
  123--140, 1996.

\bibitem{leo_breiman_bias_1996}
------, ``Bias, {Variance}, and {Arcing} {Classifiers},'' Apr. 1996.

\bibitem{zhou_ensemble_2012}
Z.-H. Zhou, \emph{Ensemble methods: foundations and algorithms}, ser. Chapman
  \& {Hall}/{CRC} machine learning \& pattern recognition series.\hskip 1em
  plus 0.5em minus 0.4em\relax Boca Raton, FL: Taylor \& Francis, 2012.

\bibitem{goodfellow_deep_2017}
I.~Goodfellow, Y.~Bengio, and A.~Courville, \emph{Deep learning}, ser. Adaptive
  computation and machine learning series.\hskip 1em plus 0.5em minus
  0.4em\relax Cambridge, MA: MIT Press, 2017.

\bibitem{jarrett_what_2009}
K.~Jarrett, K.~Kavukcuoglu, M.~Ranzato, and Y.~LeCun, ``What is the best
  multi-stage architecture for object recognition?'' in \emph{2009 {IEEE} 12th
  {International} {Conference} on {Computer} {Vision}}, Sep. 2009, pp.
  2146--2153.

\bibitem{glorot2011deep}
X.~Glorot, A.~Bordes, and Y.~Bengio, ``Deep sparse rectifier neural networks.''
  in \emph{Aistats}, vol.~15, no. 106, 2011, p. 275.

\bibitem{krizhevsky2012imagenet}
A.~Krizhevsky, I.~Sutskever, and G.~E. Hinton, ``Imagenet classification with
  deep convolutional neural networks,'' in \emph{Advances in neural information
  processing systems}, 2012, pp. 1097--1105.

\bibitem{bengio_learning_2009}
Y.~Bengio, ``\BIBforeignlanguage{en}{Learning {Deep} {Architectures} for
  {AI}},'' \emph{\BIBforeignlanguage{en}{Foundations and Trends in Machine
  Learning}}, vol.~2, no.~1, pp. 1--127, 2009.

\bibitem{hinton2006fast}
G.~E. Hinton, S.~Osindero, and Y.-W. Teh, ``A fast learning algorithm for deep
  belief nets,'' \emph{Neural computation}, vol.~18, no.~7, pp. 1527--1554,
  2006.

\bibitem{glorot2010understanding}
X.~Glorot and Y.~Bengio, ``Understanding the difficulty of training deep
  feedforward neural networks.'' in \emph{Aistats}, vol.~9, 2010, pp. 249--256.

\bibitem{he2015delving}
K.~He, X.~Zhang, S.~Ren, and J.~Sun, ``Delving deep into rectifiers: Surpassing
  human-level performance on imagenet classification,'' in \emph{Proceedings of
  the IEEE international conference on computer vision}, 2015, pp. 1026--1034.

\bibitem{mishkin_all_2016}
D.~Mishkin and J.~Matas, ``All you need is a good init,'' in \emph{{ICLR}},
  2016.

\bibitem{lecun2012efficient}
Y.~A. LeCun, L.~Bottou, G.~B. Orr, and K.-R. M{\"u}ller, ``Efficient
  backprop,'' in \emph{Neural networks: Tricks of the trade}.\hskip 1em plus
  0.5em minus 0.4em\relax Springer, 2012, pp. 9--48.

\bibitem{polyak1964some}
B.~T. Polyak, ``Some methods of speeding up the convergence of iteration
  methods,'' \emph{USSR Computational Mathematics and Mathematical Physics},
  vol.~4, no.~5, pp. 1--17, 1964.

\bibitem{duchi2011adaptive}
J.~Duchi, E.~Hazan, and Y.~Singer, ``Adaptive subgradient methods for online
  learning and stochastic optimization,'' \emph{Journal of Machine Learning
  Research}, vol.~12, no. Jul, pp. 2121--2159, 2011.

\bibitem{tieleman2012lecture}
T.~Tieleman and G.~Hinton, ``Lecture 6.5-rmsprop: Divide the gradient by a
  running average of its recent magnitude,'' \emph{COURSERA: Neural networks
  for machine learning}, vol.~4, no.~2, pp. 26--31, 2012.

\bibitem{kingma2014adam}
D.~Kingma and J.~Ba, ``Adam: A method for stochastic optimization,''
  \emph{arXiv preprint arXiv:1412.6980}, 2014.

\bibitem{srivastava2014dropout}
N.~Srivastava, G.~Hinton, A.~Krizhevsky, I.~Sutskever, and R.~Salakhutdinov,
  ``Dropout: A simple way to prevent neural networks from overfitting,''
  \emph{The Journal of Machine Learning Research}, vol.~15, no.~1, pp.
  1929--1958, 2014.

\bibitem{razavian_cnn_2014}
A.~S. Razavian, H.~Azizpour, J.~Sullivan, and S.~Carlsson, ``{CNN} {Features}
  off-the-shelf: an {Astounding} {Baseline} for {Recognition},''
  \emph{arXiv:1403.6382 [cs]}, Mar. 2014.

\bibitem{yosinski_how_2014}
J.~Yosinski, J.~Clune, Y.~Bengio, and H.~Lipson, ``How {Transferable} {Are}
  {Features} in {Deep} {Neural} {Networks}?'' in \emph{Proceedings of the 27th
  {International} {Conference} on {Neural} {Information} {Processing} {Systems}
  - {Volume} 2}, ser. {NIPS}'14.\hskip 1em plus 0.5em minus 0.4em\relax
  Cambridge, MA, USA: MIT Press, 2014, pp. 3320--3328.

\bibitem{borrelli2004channel}
A.~Borrelli \emph{et~al.}, ``Channel models for ieee 802.11 b indoor system
  design,'' in \emph{ICC'04}, vol.~6.\hskip 1em plus 0.5em minus 0.4em\relax
  IEEE, 2004, pp. 3701--3705.

\bibitem{lemic2014demo}
F.~Lemic, J.~B{\"u}sch, M.~Chwalisz, V.~Handziski, and A.~Wolisz, ``Demo
  abstract: Testbed infrastructure for benchmarking rf-based indoor
  localization solutions under controlled interference,'' in \emph{Proc. of
  11th European Conference on Wireless Sensor Networks (EWSN'14)}, 2014, pp.
  1--5.

\bibitem{lemicdata}
F.~Lemic, ``Data management services for evaluation of rf-based indoor
  localization filip lemic and vlado handziski.''

\bibitem{torres2014ujiindoorloc}
J.~Torres-Sospedra, R.~Montoliu, A.~Mart{\'\i}nez-Us{\'o}, J.~P. Avariento,
  T.~J. Arnau, M.~Benedito-Bordonau, and J.~Huerta, ``Ujiindoorloc: A new
  multi-building and multi-floor database for wlan fingerprint-based indoor
  localization problems,'' in \emph{Indoor Positioning and Indoor Navigation
  (IPIN), 2014 International Conference on}.\hskip 1em plus 0.5em minus
  0.4em\relax IEEE, 2014, pp. 261--270.

\bibitem{moreira2015wi}
A.~Moreira, M.~J. Nicolau, F.~Meneses, and A.~Costa, ``Wi-fi fingerprinting in
  the real world-rtls um at the evaal competition,'' in \emph{Indoor
  Positioning and Indoor Navigation (IPIN), 2015 International Conference
  on}.\hskip 1em plus 0.5em minus 0.4em\relax IEEE, 2015, pp. 1--10.

\end{thebibliography}
